# Australian Aboriginal Geomythology:
# Eyewitness Accounts of Cosmic Impacts?

Duane W. Hamacher[1] and Ray P. Norris[2]

## Abstract

Descriptions of cosmic impacts and meteorite falls are found throughout Australian Aboriginal oral traditions. In some cases, these texts describe the impact event in detail, suggesting that the events were witnessed, sometimes citing the location. We explore whether cosmic impacts and meteorite falls may have been witnessed by Aboriginal Australians and incorporated into their oral traditions. We discuss the complications and bias in recording and analysing oral texts but suggest that these texts may be used both to locate new impact structures or meteorites and model observed impact events. We find that, while detailed Aboriginal descriptions of cosmic impacts and meteorite falls are abundant in the literature, there is currently no physical evidence connecting any of these accounts to impact events currently known to Western science.

## Notice to Aboriginal Readers

This paper gives the names of, or references to, Aboriginal people that have passed away throughout, and to information that may be considered sacred to some groups. It also contains information published in *Nomads of the Australian Desert* by Charles P. Mountford (1976). This book was banned from sale in the Northern Territory in 1976 as it contained sacred/secret knowledge of the Pitjantjatjara (see Brown 2004:33-35; Neate 1982). No information about the Pitjantjatjara from Mountford's book is presented in this paper.

---

[1] Department of Indigenous Studies, Macquarie University, NSW, 2109, Australia
duane.hamacher@mq.edu.au | Office: (+61) 2 9850 8671

Duane Hamacher is a PhD candidate in the Department of Indigenous Studies at Macquarie University in Sydney, Australia. After graduating in physics from the University of Missouri and obtaining a master's degree in astrophysics from the University of New South Wales, he was awarded a Research Excellence Scholarship to study Aboriginal Astronomy at Macquarie. Duane is also an astronomy educator at Sydney Observatory and the Macquarie University Observatory and Planetarium.

[2] Department of Indigenous Studies, Macquarie University, NSW, 2109, Australia
ray.norris@csiro.au | Office: (+61) 2 9372 4416

Professor Ray Norris is an astrophysicist within CSIRO Astronomy & Space Science and an Adjunct Professor in the Department of Indigenous Studies at Macquarie University. While obtaining his MA and PhD in physics from Cambridge and Manchester, respectively, he began researching the archaeoastronomy of British stone arrangements. Ray is the Secretary of the International Society of Archaeoastronomy and Astronomy in Culture (ISAAC) and heads the Aboriginal Astronomy Research Group where he works with Indigenous communities such as the Wardaman and Yolngu in the Northern Territory.

## 1.0    Introduction

Australia is home to hundreds of Aboriginal groups (Walsh 1991), each with a distinct language and culture, stretching back more than 40,000 years (O'Connell and Allen 2004).  Many Australian Aboriginal cultures possess strong oral traditions and complex social systems (e.g. Maddock 1982; Ross 1986:232), including narratives and oral texts that have been handed down over many generations (Ross 1986). Threaded through these texts are accounts of geological events such as volcanic eruptions, earthquakes, and tsunamis, and descriptions of the origins of mountains and islands.  In some cases, the description indicates that these events were witnessed, resulting in a significant impact on that community, as suggested by Norman Tindale (1946:76) who stated that "Aboriginal myths may occasionally refer to some half-remembered cataclysm of nature, or an eclipse, or a meteoric shower" (Stanner 1975:5).  The study of how geological events or geographical features and materials are incorporated into oral traditions is referred to as 'Geomythology' (see Piccardi and Masse 2007).  This paper focuses on oral texts relating to meteorite falls and cosmic impacts.  Using the hypothesis that oral texts can serve as historical records of past geological events (Masse et al 2007a), we examine these records for information that could be used to locate new meteoritic sites, model meteoritic events, or measure the antiquity of dreaming stories.  Scientific data from these events, including the age, location, and impact effects, can assist in understanding the nature and evolution of oral traditions over time.

## 1.1    Aboriginal Australia: Prehistory and Orality

From a scientific perspective, the arrival of humans in Australia is a matter of contentious debate.  From approximately 70,000 to 15,000 years ago, the sea level was much lower than it is today (Voris 2001). During this period, Tasmania, Australia, and New Guinea were a single landmass, called Sahul.  It is believed that humans migrated to Sahul in multiple waves from South East Asia, though it seems Aboriginal Australians share a common genetic thread with people from southern India as opposed to Indonesia or Malaysia (Redd and Stoneking 1999).  The exact date of human arrival to Sahul is uncertain, but solid archaeological evidence suggests that Australia has been inhabited for at least the last 45,000 years (O'Connell and Allen 2004).  Some archaeological sites suggest an earlier arrival date, such as 60,000 to 70,000 years at Lake Mungo in New South Wales (Adcock et al 2001; Thorne et al 1999) and 60,000 years at Nauwalabila I in the Northern Territory (Roberts et al 1999).  However, further analyses of these sites have constrained the age to between 40,000 and 50,000 years (Bowler et al 2003; Gillespie and Roberts 2000; Bird et al 2002).  It is generally accepted by mainstream archaeologists that the age of human habitation of Australia is *at least* 40,000 years.

From the perspective of many Aboriginal communities, Aboriginal Australians have *always* been in Australia, and their cosmologies describe the origins of their ancestors and traditions in various forms.  These cosmologies are recorded in the oral tradition, which itself is passed down subsequent generations, evolving as new events are incorporated into art, song and storyline.  These oral texts preserve the knowledge, moral codes, laws, and traditions necessary for the survival, cohesion, and social structure of the community.

> *"It is very difficult, as things stand, for anyone, scholar or layman, to gain an overall impression of the nature and extent of Aboriginal oral tradition"* (Ross 1986:260).

Australian Aboriginal oral traditions are typically considered components of the 'Dreaming', a term coined by anthropologist Francis James Gillen in 1896 to refer to the period in the religious mythologies of northern Arunta people of the Northern Territory (Dean 1996). The Dreaming is a highly complex concept that exhibits substantial diversity in meaning and purpose between different Aboriginal communities and cannot be simply regarded or dismissed as mythology or fables. We acknowledge that the Indigenous Australian and the Western scientific methods and approach to acquiring knowledge and explaining the natural world are vastly different. We seek only to apply information from oral traditions to the discipline of geomythology. This work is in no way meant to "legitimise" or "marginalise" Aboriginal oral traditions or cultures. For a better understanding of the Dreaming, we refer the interested reader to Stanner (1958, 1965, 1975), Meggitt (1972), Charlesworth et al (1984), Ross (1986), Rumsey (1994), Beckett (1994), Dean (1996), Bates (1996), and Rose (1996, 2000).

Given the substantial diversity of Aboriginal communities, customs, laws, and traditions may vary between Aboriginal communities, even those of the same language group. While some common themes exist across many Aboriginal communities, such as the concept of the Rainbow Serpent (Radcliffe-Brown 1926), each group has a different variation of these themes. For this reason, each reference to a particular Aboriginal story, ethnography, or word in this paper will include the name and location of that particular Aboriginal group. A full list of the Aboriginal groups discussed in this paper, including their location, can be found in the Appendix.

In many Aboriginal communities, certain ceremonies and oral traditions are considered sacred and secret, and thus are not shared with Westerners or even non-initiated members of that community. For example, Barker (1964:109-110) tells how a group of Aboriginal people knew of a meteorite in the desert and described how it fell to the earth from the sky, leaving little doubt in his mind that his Aboriginal informants had witnessed the fall. But since the meteorite was used as a source for sacred stories, the informants would not reveal its location to Barker or the other colonists. It is possible that some of the conclusions reached in this paper, particularly regarding the lack of stories about known impact craters, may be affected by this need for secrecy.

## 1.2    Australian Impact Craters

Australia is home to 27 confirmed terrestrial impact structures and numerous suspected terrestrial and submarine structures (e.g. Gibbons 1977; Earth Impact Database 2009; Becker et al 2004; Abbott et al 2005a; Green 2008; Martos et al 2006). These structures range in size from the 20 meter Dalgaranga crater to the 90 km Acraman crater, and vary in age from billions of years to a few thousand years. Table 1a lists all confirmed Australian craters, while Table 1b lists all probable or suspected craters, shown in Figure

1.   The vast majority (21 out of 27) of impact structures are found in the Northern Territory and Western Australia.   There are only two confirmed in Queensland, one suspected crater in Tasmania, and no confirmed craters in New South Wales or Victoria. The recent discovery of a possible impact structure in northwest New South Wales (Green 2008) may be that state's first known crater.

Of the confirmed craters in Australia, the constrained dates of only three occur within well-established human habitation of the continent (< 40,000 years).   These are the Henbury crater field and Boxhole crater in the Central Desert and the Veevers crater in Western Australia.

**Table 1a:**  Confirmed terrestrial impact craters in Australia based on data from Earth Impact Database (2009), numbered (N) as in Figure 1a.  States are abbreviated as follows: Western Australia = WA, Northern Territory = NT, South Australia = SA, Queensland = QLD, New South Wales = NSW (including the Australian Capital Territory), Victoria = VIC, and Tasmania = TAS.  The estimated crater age is given in millions of years (Ma) and the crater diameter (D) is given in km.  Craters with an established age consistent with human habitation of Australia are noted with an asterix (*).

| N | Name | Latitude | Longitude | State | Age (Ma) | D (km) |
|---|------|----------|-----------|-------|----------|--------|
| 1 | Woodleigh | 26° 03′ S | 114° 39′ E | WA | 364±8 | 40 |
| 2 | Dalgaranga | 27° 38′ S | 117° 17′ E | WA | ~ 0.27 | 0.02 |
| 3 | Yarrabubba | 27° 10′ S | 118° 50′ E | WA | ~ 2000 | 30 |
| 4 | Shoemaker (Teague) | 25° 52′ S | 120° 53′ E | WA | 1630±5 | 30 |
| 5 | Glikson | 23° 59′ S | 121° 34′ E | WA | < 508 | ~19 |
| 6 | Connolly Basin | 23° 32′ S | 124° 45′ E | WA | < 60 | 9 |
| 7 | Veevers* | 22° 58′ S | 125° 22′ E | WA | < 0.02 | 0.08 |
| 8 | Wolfe Creek | 19° 10′ S | 127° 48′ E | WA | < 0.3 | 0.87 |
| 9 | Goat Paddock | 18° 20′ S | 126° 40′ E | WA | < 50 | 5.1 |
| 10 | Spider | 16° 44′ S | 126° 05′ E | WA | > 570 | 13 |
| 11 | Piccaninny | 17° 32′ S | 128° 25′ E | WA | < 360 | 7 |
| 12 | Matt Wilson[a] | 15° 27′ S | 130° 28′ E | NT | unknown | 7.5 |
| 13 | Liverpool | 12° 24′ S | 134° 03′ E | NT | 150±70 | 1.6 |
| 14 | Goyder | 13° 28′ S | 135° 02′ E | NT | < 1400 | 3 |
| 15 | Strangeways | 15° 12′ S | 133° 35′ E | NT | 646±42 | 25 |
| 16 | Foelsche | 16° 40′ S | 136° 47′ E | NT | > 545 | 6 |
| 17 | Lawn Hill | 18° 40′ S | 138° 39′ E | QLD | > 515 | 18 |
| 18 | Kelly West | 19° 56′ S | 133° 57′ E | NT | > 550 | 10 |
| 19 | Amelia Creek | 20° 55′ S | 134° 50′ E | NT | 600 to 1640 | 90 |
| 20 | Boxhole* | 22° 37′ S | 135° 12′ E | NT | 0.0054±0.0015 | 0.17 |
| 21 | Gosses Bluff | 23° 49′ S | 132° 19′ E | NT | 142.5±0.8 | 22 |
| 22 | Henbury* [b] | 24° 34′ S | 133° 08′ E | NT | 0.0042±0.0019 | 0.15 |
| 23 | Tookoonooka | 27° 07′ S | 142° 50′ E | QLD | 128±5 | 55 |
| 24 | Mount Toondina | 27° 57′ S | 135° 22′ E | SA | < 110 | 4 |
| 25 | Acraman | 32° 01′ S | 135° 27′ E | SA | ~ 590 | 90 |
| 26 | Flaxman | 34° 37′ S | 139° 04′ E | SA | > 35 | 10 |
| 27 | Crawford | 34° 43′ S | 139° 02′ E | SA | > 35 | 8.5 |

[a] Kenkmann and Poelchau (2008)

[b] The Henbury crater field comprises 13 craters, resulting from the break-up of a nickel-iron meteoroid in the atmosphere before impact.  The diameter given in Table 1a is of the largest crater, which is actually a cluster of three craters superimposed.

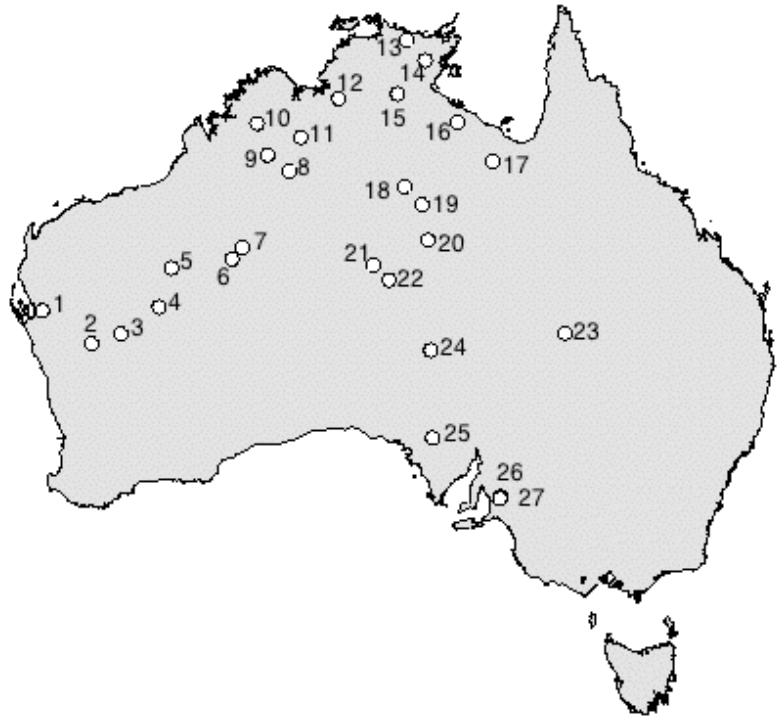

**Figure 1a:** Geographic locations of known terrestrial impact craters in Australia. Craters are numbered from Table 1a. The Flaxman and Crawford craters near Adelaide, South Australia (26 and 27) are very close in proximity and appear almost completely superimposed on this map.

**Table 1b:** Probable or suspected terrestrial impact structures in Australia, numbered (N) as in Figure 1b. Wulinggi is the local Aboriginal name for Crater Lake near Batchelor, considered a sacred site of the Kungarakany and Warai people (Murgatroyd 2001:13).

| N | Name | Latitude | Longitude | State | Reference |
|---|------|----------|-----------|-------|-----------|
| 1 | Hickman | 23° 02′ S | 119° 40′ E | WA | Glickson et al (2008) |
| 2 | Wulinggi | 13° 02′ S | 131° 06′ E | NT | Murgatroyd (2001) |
| 3 | Maningrida | 11° 53' S | 134° 12' E | NT | Haines (2005) |
| 4 | Gulpuliyul | 13° 19′ S | 134° 06' E | NT | Sweet et al (1999) |
| 5 | Renehan | 18° 18' S | 132° 39' E | NT | Haines (2005) |
| 6 | Spear Creek | 17° 16' S | 135° 50' E | NT | Kruse et al (2010:59) |
| 7 | Calvert Hills | 17° 22' S | 137° 28' E | SA | Macdonald and Mitchell (2004) |
| 8 | Wessel | 20° 36' S | 135° 05' E | NT | Macdonald and Mitchell (2004) |
| 9 | White Cliffs | 30° 47' S | 143° 08′ E | NSW | Green (2008) |
| 10 | Darwin | 42° 18′ S | 145° 39' E | TAS | Howard and Haines (2007) |

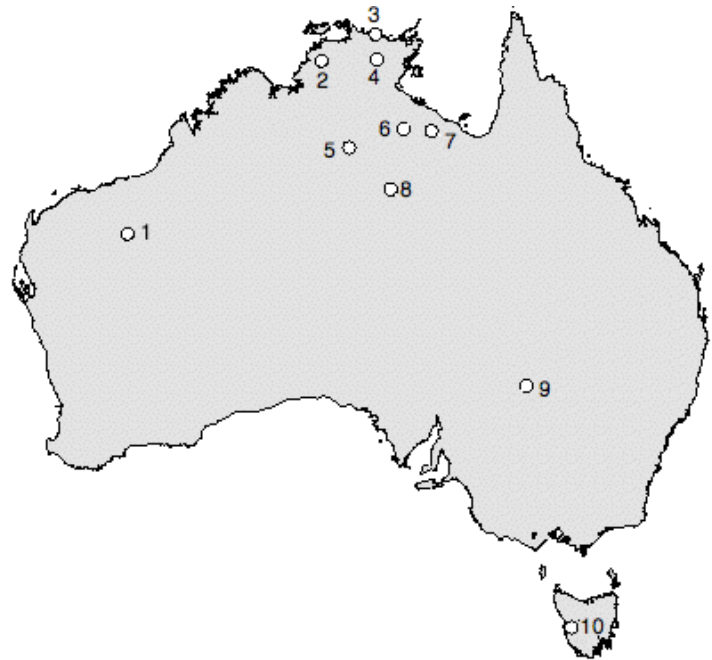

**Figure 1b:** Geographic locations of probable or suspected terrestrial impact craters in Australia. Craters are numbered from Table 1b.

O'Connell (1965) mentions a handful of crater candidates in Australia that are not listed above, including a crater "seen on an Aboriginal reserve on Koolatong River near Blue Mud Bay" (13° 10' S, 135° 40' E, *ibid*:16), the Lake Hamilton craters on the west coast of Eyre Peninsula (SA 33° 55' S, 135° 16' E, *ibid*:70) and the Weepra Park Depressions 30 km east of Elliston, Eyre Peninsula, SA (33° 26' S, 134° 16' E, *ibid*:125) which are both estimated to be < 9,000 years old and are probably solution-collapsed domes. She also includes the Mount Doreen crater, NT at the general coordinates of 23° S, 133° E which is estimated to have formed ~100 years ago (*ibid*:88).

## 1.3    Geomythology

The study of geological phenomena (including astronomical events such as comets, eclipses and supernovae) inspiring aspects of oral traditions is referred to as 'Geomythology', a term coined in 1968 by Dorothy Vitaliano, a geologist at Indiana University (Vitaliano 1968). The word "mythology" is derived from the Greek "mythos", meaning 'story' (or in some cases 'word') and "logos", meaning 'word' in the form of speech. Thus, mythology can be considered "spoken stories". There is a long-standing debate regarding the nature and purpose of myth, and associated theories of myth vary significantly between academic disciplines. Here, we draw from Masse et al (2007a) that myths can serve as oral records of natural events and that these records can be elicited by modern science to understand or model historic natural events – in this case, cosmic impact events[3]. During the twentieth century, the use of myth to explain geological events was viewed with scepticism by the scientific community, as there was little physical evidence to support the hypothesis. However, in the 21[st] century, physical evidence is helping geomythology gain new ground as a legitimate discipline. The first major peer-reviewed work on geomythology is "Myth and Geology" (Piccardi and Masse 2007), which provides the theoretical foundation of the discipline and serves as the methodological basis of this work. Given the generally negative connotations of the word "mythology", we instead use the more accurate term "oral tradition".

---

[3] For a similar study applied to South America, see Masse and Masse (2007).

While Piccardi and Masse (2007) provide a plethora of examples of geomythology from around the world, they represent only the tip of the iceberg. Many stories use natural events or landforms to illustrate a moral point, sometimes including warnings to future generations. On 26 December 2004, such an example was witnessed by the world as an earthquake off the western coast of Sumatra induced a tsunami that swept across the Indian Ocean, killing thousands. Some indigenous peoples of the Indian Ocean survived the tsunami because of information contained in their mythology. These groups, including the Moken people of Thailand and Indigenous Andaman Islanders, possessed stories that told of great waves that would "eat men" and that this wave would come when the sea receded rapidly. Their stories told them that to survive they must immediately run to high ground. Their adherence to these stories saved their lives (Arunotai 2006; Masse et al 2007a:18). This suggests that such events probably occurred in the past (see Jankaew et al 2008) and were significant enough to be incorporated into story lines that lasted long periods of time, and these stories contained information and warnings about natural catastrophes that were crucial to the survival of the community.

One of the most well-documented examples of geomythology in Australia are the stories describing the volcanic eruptions that formed the Eacham, Barrine, and Euramo crater lakes in Queensland, which formed over 10,000 years ago (Dixon 1972:29; Rainforest Conservation Society of Queensland 1986:39). The stories describe the region as covered in Eucalypt scrub as opposed to the current rainforest. This was later confirmed by the analysis of fossil pollen found in the silt of these craters, which showed the current rainforest to be 7,600 years old (Kershaw 1970; Haberle 2005). The Australian Heritage Commission includes these stories on the Register of the National Estate and within Australia's World Heritage nomination of the wet tropical forests as an "unparalleled human record of events dating back to the Pleistocene era" (Pannell 2006:11; Rainforest Conservation Society of Queensland 1986:40).

Similarly, Native American stories describe the eruption of Mount Mazama, which formed Crater Lake in Oregon. Sandals excavated under Mazama ash have been radiocarbon dated at 6,500 years old (Vitaliano 2007), showing the area was inhabited and affected at the time of eruption. Stories from South America suggest Indigenous cultures not only witnessed volcanic eruptions, but that the descriptions of these events remained in the oral tradition for hundreds, perhaps thousands, of years (Masse and Masse 2007).

Another example of geomythology from Australia and New Zealand is given by Bryant (2001, et al 2007), who suggests that the southeast coast of Australia was struck by a tsunami within the last 600 years. Bryant proposes that the tsunami was induced by a cosmic impact in the Tasman Sea and cites various Aboriginal and Maori stories describing tsunamis or cosmic impacts followed by a deluge to support his claim, including the Maori story about the "Fires of Tamatea". This view was challenged by Goff et al (2003), who argued that the Maori placenames were mistranslated and that no 'smoking gun' (crater) existed. However, a submarine structure, named Mahuika after the Maori god of fire, was discovered in the Tasman Sea south of New Zealand by Abbott et al (2003) and may be an impact crater 20±2 km in diameter with an estimated impact date of ~1443 CE (Abbott et al 2005a), supporting Bryant's hypothesis. However, the dating of this structure and its origins are still the topic of contentious debate.

## 2.0    Known Australian Impact Craters

### 2.1    Gosse's Bluff

Gosse's Bluff, approximately 200 km west of Alice Springs, is an eroded impact crater with a diameter of ~22 km and a remnant circular uplift forming a mountain range ~5 km in diameter and ~150 m in height with an age of 142.5±0.8 Ma (Milton et al 1996). Scientists first proposed impact origins of the structure in the 1960s based on the abundance of shatter cones (Dietz 1967). To the Western Arrernte[4] people, Gosse's Bluff is known as Tnorala and is considered a sacred place. An Arrernte story regarding its origins closely parallels the scientific explanation: in the Dreaming, a group of sky-women were dancing as stars in the Milky Way. One of the women grew tired and placed her baby in a wooden basket, called a turna. As the women continued dancing, the turna fell and plunged into the earth. The baby fell and was covered by the turna, which forced the rocks upward, forming the circular mountain range. The baby's mother, the evening star, and father, the morning star, continue to search for their baby to this day (Parks and Wildlife Commission of the Northern Territory 1997:1; Cauchi 2003; Williams 2004; Malbunka 2009).

### 2.2    Wolfe Creek

Wolfe Creek crater, located in northeastern Western Australia, is 850x900 m in diameter with an estimated age of ~300,000 years (Shoemaker et al 1990; 2005). The area is home to the Djaru, who call the structure Kandimalal (Tindale 2005:376). According to Cassidy (1954:198), Kandimalal has "no particular meaning in their language, and no legend exists to give a hint of its origin." However, the literature reveals multiple stories associated with the crater, some of which describe cosmic origins (see Sanday 2007). One of the earliest Djaru accounts tells how a pair of subterranean Rainbow Serpents created the nearby Wolfe and Sturt creeks. One serpent emerged from the ground, creating the circular structure (Bevan and McNamara 1993:6; Goldsmith 2000). Another story tells how one night, the moon and the evening star passed very close to each other. The evening star became very hot and fell to the earth, causing a brilliant, deafening explosion. This greatly frightened the Djaru and it was a long time before they ventured near the site, only to discover it was the spot where the evening star had fallen (Goldsmith 2000). Goldsmith reports that the Aboriginal Elder told him this story came from his grandfather's grandfather, indicating it was handed down in its present form before the scientific identification of the crater. A Djaru Elder named Jack Jugarie (1927-1999) gave his account of the Wolfe Creek crater: "A star bin fall down. It was a small star, not so big. It fell straight down and hit the ground. It fell straight down and made that hole round, a very deep hole. The earth shook when that star fell down" (Sanday 2007:26). Speiler Sturt, a Djaru Elder from Billiluna, Western Australia, also illustrates the cosmic origins of Wolfe Creek crater (*ibid*, see Figure 2):

> *That star is a Rainbow Serpent*
> *This is the Aboriginal Way*
> *We call that snake Warnayarra*
> *That snake travels like stars travel in the sky*
> *It came down at Kandimalal (Wolfe Creek)*
> *I been there, I still look for that crater*
> *I gottem Ngurriny – that one, Walmajarri/Djaru wild man.*

---

[4] Arrernte, Aranda, and Arunta are all different spellings of the same Aboriginal language group of Central Australia. Different spellings are used in this paper according to the spelling given in the source from which the account was taken.

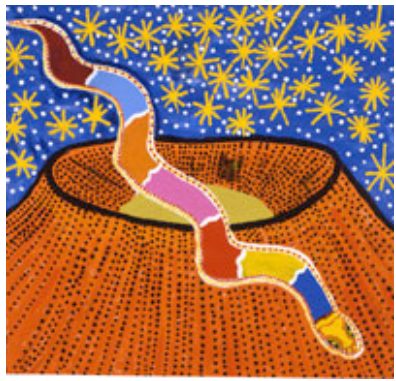

**Figure 2:** Wolfe Creek Crater and the Rainbow Serpent. Painting (2000) by Boxer Milner, a Djaru Elder from Billiluna, Western Australia. Image reproduced with permission, courtesy of Peggy Reeves Sanday (2007), http://www.sas.upenn.edu/~psanday/Aboriginal/serpent4.html.

## 2.3 Henbury Crater

With an age of 4200±1900 years, the Henbury impact event was probably witnessed firsthand since people have inhabited the Central Desert for more than 20,000 years (see Smith 1987; Smith et al 2001). However, Alderman (1931:28) concludes that the local Aboriginal people seemed to have "no interest" in the Henbury craters or ideas as to their origins. An addendum by L.J. Spencer in Alderman (1931:31), however, suggests that Aboriginal people viewed the Henbury impacts with apprehension. A local prospector named J.M. Mitchell noted to Spencer that older Aboriginal people would not camp within a couple of miles of the Henbury craters and referred to them as "chindu china waru chingi yabu", roughly translating to "sun walk fire devil rock" (*ibid*:31). Fitzgerald (1979) claims that Aboriginal people would not drink the water from the craters in fear that the "fire devil" would fill them with iron. These views and the name suggest recognition of the site as different from the surrounding landscape, highlighting the iron fragments that litter the area and the site's unusual nature. However, not all Aboriginal people of the area shared this apprehension. Brown (1975:190-191) notes that the Henbury craters were an important water source to the local Aboriginal people, as the craters collected and retained rainwater for long periods of time. Although a story was recorded by Mountford (1976:259-260) describing the crater's origins, it did not attribute the structure's formation to a cosmic impact. According to the Parks and Wildlife Commission of the Northern Territory (2002:15), the Arrernte name for the crater field is Tatyeye Kepmwere (Tatjakapara) and that "some of the mythologies [sic] for the area are known but will only be used for interpretation purposes after agreement by the Aboriginal custodians of the site".

## 2.4    Veevers Crater

In the center of Western Australia lies the small, 70 meter-wide Veevers crater, considered one of the best preserved small impact craters in the world and one of a minority that is associated with meteorite fragments (Shoemaker et al 2005). Using cosmogenic nuclide exposure of the crater walls, the estimated age of Veevers crater is less than 20,000 years old. However, the well-preserved ejecta suggest it may be less than 4,000 years old (*ibid*). In either case, the Veevers impact may have been witnessed by humans, but no Aboriginal stories about it are recorded in the literature.

## 2.5    Boxhole Crater

Boxhole crater is a 170-m wide impact structure located ~170 km northeast of Alice Springs (see map in Figure 1). The age of this structure is contentious, as Kohman and Goel (1963) report an age of 5,400±1,500 years using $^{14}$C exposure, while Shoemaker et al (1990; 2005) report an age of ~30,000 years using $^{10}$Be/$^{26}$Al exposure (see Haines 2005:484-485). In either case, the impact may have been witnessed, depending on the precise date of human arrival to the Boxhole region. Madigan (1937:190) claimed that the local Aboriginal people seemed to have "no interest" in the crater and nothing else is reported in the literature.

## 2.6    Other craters

A search for stories associated with other confirmed impact craters revealed a story about Liverpool crater in Arnhem Land (Haines 2005), which is described as the nest of a giant catfish. This story is supported by pictographs of giant catfish on the walls of a rock shelter located within the crater (Shoemaker and MacDonald 2005). The crater is ~1.6 km wide and dates to ~150 Ma. Except for Gosse's Bluff, Wolfe Creek, Liverpool and Henbury, we were unable to locate associated stories relating to other confirmed impact craters.

## 2.7    Summary of the evidence from known impact craters

### 2.7.1    Why are there stories of Wolfe Creek and Gosse's Bluff?

Two Australian impact craters, Gosse's Bluff and Wolfe Creek, have associated stories that attribute extraterrestrial origins to the structures' formation, even though the craters were formed long before human habitation of Australia. It is unclear if (a) the stories describing Gosse's Bluff and Wolfe Creek as cosmic impacts incorporated the modern scientific explanation, (b) the Aboriginal people had deduced, without the influence of Western science, that these structures were formed by cosmic impacts, (c) parallels between scientific and Aboriginal views are simply coincidence, or d) the information the Aboriginal elders provided to the researchers was misleading.

Misleading information can be caused by a poorly worded question (e.g. Researcher: "Do you have any stories about that meteorite crater we call […]?"), in which case the Aboriginal informant may tell the researcher what they "want to hear" out of courtesy or respect (e.g. Clarkson 1999). The first argument (a) could be falsified by showing, conclusively, that the cosmic impact stories predate the scientific discovery of the craters. Addressing (b), (c), and (d) requires a more detailed understanding of Aboriginal knowledge systems, Dreamings, and Aboriginal customs.

### 2.7.2   Why are there no stories of the known impact craters that occurred in human history?

Three craters, Henbury, Boxhole and Veevers, were formed during human habitation of Australia and would undoubtedly have caused destruction in the region of the impact site. Although there are some stories related to the Henbury craters (although some are not in the public domain) it is curious that no story exists regarding Veevers or Boxhole. Possible explanations are that (a) humans did not inhabit that area at that time, (b) that stories do exist, but are secret and thus not revealed to an uninitiated researcher, or (c) that stories did once exist, but have been lost in time, perhaps through cultural discontinuity. We do not currently have the information necessary to discriminate between these explanations.

### 3.0   Other Impact Sites

A survey of the available literature revealed a plethora of stories and accounts that describe stars falling from the sky and crashing to the earth. These stories, which do not correspond to impact structures or meteorite finds known to Western science, could be used to model possible impact catastrophes (e.g. Masse and Masse 2007) or assist in locating new impact sites. Some stories are somewhat vague and do not cite a specific location, such as that from the Wolmeri of the Kimberleys who tell how Venus came to earth and left a stone in one of the "horde countries" (Kaberry 1939:12), a Wawilak (Yolngu) place in Arnhem Land called "Katatanga", meaning "the falling meteor place" (Warner 1937:251), or the Indjabinda story about Mangela, who came to the earth with a fleeting appearance at a place denoted by a sacred mound called Kumana Kira (Cowan 1992:28). Other stories, however, cite the specific location of the event. Here, we sort the stories by state, citing locations where available. The coordinates of all locations with a superscript '*f*' are given in Table 2 and shown in Figure 3. To facilitate later discussion, those events that may correspond to a historical event are assigned an "Event Number".

## 3.1    New South Wales

Several oral traditions from New South Wales (NSW) describe cosmic impacts, especially in the northwest region of the state. A story from the Muruwari of north-central NSW describes a catastrophic event that Mathews (1994:60) interprets to be a meteorite impact, citing a large circular area near the village of Bodena*f*, where Mathews indicates the event took place (Event #1). The story, recorded between 1968 and 1972, involves fire falling from the sky to the earth, causing death and destruction. Gien, the man responsible for the catastrophe, fled to the sky and became the moon. Mathews recounts a nearly identical story from the nearby Ngemba peoples, who tell of a fireball that landed on the camp, killing everyone. A similar story from the nearby Kula peoples (recorded in 1804, *ibid*) describes a giant piece of burning wood that fell upon the camp, with the responsible party, Giwa, becoming the moon. The similar stories probably represent variations of a single account that has been incorporated into the storylines of adjacent Aboriginal communities, with each group reciting their own interpretation of the story. While Mathews suggests the event took place near Bodena, the Kula story cites the location as Multaguna Run near the Warrego River, approximately 90 km west of Bodena. It is uncertain where the story originated, but the earliest record in the literature is the Kula account. No geophysical survey of the site confirming or rejecting this hypothesis is reported in the available literature.

A Weilan story (McKay 2001:112-114) tells of a female shape-shifter who would lure men away from their camps and carry them away. A man cunningly led her to a water hole, where he tied a cord around her (made of human hair) while she slept. She awoke and fought to get him off her back, jumping into the water turning it a dull colour. He speared her in the back, but she would not die. She flew up to the sky with him still clinging on until they were in the path of falling stars. A falling star knocked the man off and he fell with some stars to the ground (Event #2) at a site known today as the town of Girilambone*f*. The Weilan (*ibid*) also tell of a large star that fell to earth, lighting up all of the surrounding land - so bright, in fact, that several different Aboriginal groups saw it (e.g. Murrawarri, Barkinji, Weilan, Ngemba, Narran, Kamilaroi, and Wongaibon). A nearly identical story from the Ooungyee people of the Kimberleys in Western Australia is given in Sawtell (1955:21-22). The Sawtell account was published 46 years prior to the NSW account, though the story's origins are unclear. The Weilan story was published in a magazine "for the Aboriginal people of New South Wales", suggesting the story may have been adopted later by the Weilan. Given the nearly identical wording and theme of the text, we do not consider these to be independently contrived stories. There are no known impact craters in New South Wales, except for the newly discovered, albeit unconfirmed, structure near the town of White Cliffs, which is estimated to be more than two million years old (Macey 2008, see Figure 1b, Table 1b).

According to Threlkeld (1834:51), recounted in Turbet (1989:126), the scattered rocks at a place called Kurra Kurran*f*, on Fennell Bay at the northwestern extremity of Lake

Macquarie, between Sydney and Newcastle, lies an area of petrified wood. According to the local Awabakal people, the fragments are remnants of a large rock that was cast from the sky by a giant goanna, killing many people (Event #3). Jones (1989) recounts a story of a large fiery star that rumbled and smoked as it fell from the sky, crashing into the ground near Wilcannia[f] (Event #4). The impact was followed by a deluge, which Jones suggests is represented in rock paintings on Mount Grenfell, northwest of Cobar. The paintings show people climbing the mountain, supposedly to escape the rising water levels. Jones noted the presence of unusual black stones in the riverbed, indirectly suggesting they were meteorites. The story gives the impact location as the Darling riverbed northeast of Wilcannia at a place called *Purli Ngaangkalitji*. A survey of the area by Scottish-Australian astronomer Robert McNaught found no evidence of an impact crater or meteoritic material (Steel and Snow 1992:572). Bevan and Bindon (1996:95) suggest that the story was recounted at this spot on the Darling River, but was not where the event actually took place. If the story describes a real event, the location of the original site is currently unknown to Westerners.

Stories from the Shoalhaven region near Nowra describe an impacting meteor shower and airburst (Peck 1933:192-193, Event #5):

> "The sky moved, heaved and billowed. The stars tumbled and clattered and fell against one another. The Milky Way split and great star groups were scattered. Many stars fell to the earth, flashing in the sky. A large red glowing mass burst in the air, giving a deafening roar, as it scattered millions of molten pieces on the ground. This occurred all night. The ground was covered in burned holes and great mounds, formed from the falling pieces".

A similar story is recounted in Peck (1925:152-160); during a battle between two groups, a bright, burning blue light came hissing toward the earth from the sky. Surprised by this sight, the people thought it must have been an unknown great sorcerer. The earth trembled as the star crashed into the ground, sending rock and debris into the air. The noise of the explosion was deafening and reverberated around the surrounding hills. The men were filled with terror and they all fell flat to the ground (Event #6). Peck vaguely says this area is in "the belt of basalt country where the waratah does not grow" (*ibid*:152).

## 3.2    Northern Territory

Two stories of falling stars in Arnhem Land describe fire resulting from the impact. A Gurudara story describes a bright star[5], named Nyimibili, falling from the sky onto the Marabibi[f] camp near the Wildman River, which burned all of the grass and trees, causing death and destruction (Berndt and Berndt 1989:25-27, Event #7). The Yolngu tell of Goorda, a fire-spirit from the Southern Cross, who came to the earth as a falling star to bring fire to people of the Gainmaui River, near Caledon Bay[f]. Once he touched the ground (Event #8), he set the grass ablaze, which spread causing death and chaos (Allen 1975:109)[6].

---

[5] The fiery star, Nyimibili, was described as having a large eye as bright as the moon, burning alight during the day and night. This description closely relates to other malevolent, one-eyed meteor beings found in the oral traditions of Aboriginal people of the upper Northern Territory, such as Namorrodor and Papinjuwari.

Many Aboriginal communities believe natural disasters are punishment for breaking laws and traditions. The Wardaman people warn of a fiery star named Utdjungon that will fall from the sky to destroy the earth if laws and traditions are not followed. The falling star will cause the earth to shudder, the hills and trees to topple and turn, and everything going black with night descending (Harney and Elkin 1949:29-31). A nearly identical story from the nearby Ngarinman tells of "a large black stone, thrown from the sky by Utdjungon – a killer who lives in the Milky Way and flings a fiery ball to slay" (Harney and Elkin 1949:72-74, Event #9). No specific locations are cited in these stories, but presumably take place within Wardaman or Ngarinman country.

Some stories of falling stars have Western religious overtones. In the town of Hermannsburg[f] (Ntaria), ~110 km west of Alice Springs, an Arrernte woman told of a star that fell to earth (Event #10) creating two holes on the site of the old Hermannsburg church prior to settlement of the area by Lutheran missionaries (Austin-Broos 1994:142; 2009:39). The story was heavily laced with Christian influence and symbolism, resulting from the close association between the Aboriginal community and the missionaries. Austin-Broos (1994:149; 2009:37-38) believes this account may relate to the story of a star that fell to earth in a waterhole in nearby Palm Valley called *Puka*[f]. Róheim (1945:183) provides an earlier record of this story, citing Western and Eastern Arrernte folklore about a star that fell into a waterhole where the serpent, Kulaia, lived, making a great noise like thunder. In the story, a thirsty boy who had only recently been circumcised (initiated) peered into Puka and saw the serpent, which rose and swallowed him whole. The boy's death caused much grief and mourning among his community, who promptly burnt the food they had collected for the boy's initiation and left the camp.

In some cases, the progenitors of knowledge or humankind were brought to earth via falling stars. The Western Arrernte say the first human couple originated from a pair of stones that were thrown from the sky by the spirit Arbmaburinga (Róheim 1971:370). Yarrungkanyi and Warlpiri of the Northern Territory tell how celestial Dreaming men fell to the earth as falling stars, bringing the Dreaming to the people. The Dreaming men, armed with weapons, travelled through the sky and landed at a place called *Purrparlarla* (Warlukurlangu Artists 1987:127), southwest of Yuendumu[f] (aka Yurntumu, *ibid*:4). Another description of a "sky man" falling to the earth is from Bickerton Island[f], between Groote Eylandt and the mainland. Connie Bush (Pascoe 1990:31-32) recounts a story (Event #11) about a circular spot in the sand that was formed when a "terrific explosion in the sky" towards the west dropped a smoking white object to the earth early in the morning. At the supposed impact site, two Aboriginal men saw a white man "standing in the sand up to his knees" about an hour after the impact. The man lived with the community for many years, marrying one of the women and having a child until he finally died. The story notes that the white man's son died as a young man from a measles epidemic that spread to the island from Groote Eylandt before the arrival of missionaries in 1921. It may be related to a measles epidemic on Bathurst Island in 1913, but it is not clear. This implies that the event occurred during the end of the 19th century, ruling out a plane crash as the cause. According to Bush, the circle is still void of vegetation and shells are placed around the perimeter to remind people where the sky man fell. However, the discovery of the man and the bolide event may be purely coincidental, or perhaps the two events were incorporated into a single storyline.

---

[6] There is some debate as to whether meteorite impacts can trigger fires (e.g. Jones and Lim 2000; Durda and Kring 2004). Masse and Masse (2007:196) cite several examples from South America describing fires that were caused by fiery objects falling from the sky. Collins et al (2005) cite the heat-energy levels required to ignite grass and trees, among other materials, based on nuclear tests by Glasstone and Dolan (1977).

### 3.3 Western and South Australia

In Western Australia, Wirrimanu artists tell a story of a star (Event #12) that fell from the sky to Lake Mackay[f], explaining that the "Rainbow snake came to this place and ate up a lot of people" (Bevan and Bindon 1996:96). No meteorite finds or impact sites are currently known in or near Lake Mackay. Another Dreaming tells of a hunter named Mangowa and how his actions led to a group of stars falling to the earth (Event #13), forming the circular lagoons that line the coast of the Nullarbor (e.g. Roberts and Mountford 1965:52; Smith 1970:24; Reed 1993:237-239). A Wirangu story describes a meteorite impact (Event #14) near the coastal town of Eucla[f], close to the border of South and Western Australia (Education Department of South Australia 1992:32-33, retold by M. Miller and W. J. Miller):

> "A long, long time ago, a huge meteorite hurtled towards the earth from the northward sky, and smashed into the ground near Eucla. Because it was so big, a dent appeared in the crust of the earth and the meteorite bounced high into the air and out into the Great Australian Bight where it landed with an enormous sizzling splash. It was hot from its trip through space so it gave off a great deal of steam and gas as it sank through the waves. But this was no ordinary meteorite. It fact, it was the spirit Tjugud. In the deep water near by, the spirit woman Tjuguda lay asleep. All the noise around her woke her up and she was very angry. She bellowed and the elements roared with her. The wind blew, the rain pelted from the sky and the dust swirled."

Because of the generous use of Western scientific terms in the story ("meteorite", "crust of the earth", etc), it is uncertain how close the story is to its original form, assuming it is pre-colonisation. There are no known impact craters on the Nullarbor, but Gibbons (1977:265) and Bevan and Binns (1989) reveal numerous meteorites that have been found in area, including those found near the town of Forrest[f], ~100 km northwest of Eucla.

A story (Event #15) associating a fiery catastrophe with a large circular depression on the McGrath Flat[f] (aka Magrath Flat) homestead in South Australia prompted Tindale (1938:18) to conclude that the story described a meteorite impact (Stanner 1975:14). To date, no meteoritic material has been found in or near the structure, but it is uncertain if the structure has been properly surveyed for evidence of an impact.

When Ngalia men shared sacred information about the Walanari - celestial deities who were seen as protectors of good men and punishers of bad men that lived in the Magellanic Clouds - with the anthropologist Charles P. Mountford (1976:457), the Walanari became very angry and threw glowing stones onto the Ngalia camp later that night. Ngalia informants claim that men have been killed by falling stars thrown by the Walanari (Event #16). They said that when totemic ceremonies were being performed at Mount Doreen[f], the Walanari threw meteorites to the earth to express their pleasure (see *ibid*, Plate 584). O'Connell (1965:88) mentions a crater candidate near Mount Doreen with general coordinates of 23 S, 133 E that may have impacted ~100 years ago and states that a "local legend of fiery snake may refer to meteorite fall", but gives no references to the legend. Meteors are often associated with serpents (see Hamacher and Norris 2010b).

The Kimberleys are rich in stories of objects falling from the sky, as well as floods that indicate tsunamis. Bryant (2001, et al 2007) proposes a link between these stories, hypothesizing that cosmic debris impacted the Indian Ocean creating tsunamis that devastated the coastline. For example, Bryant (*ibid*) cites published evidence of mega-tsunami on the northwest coast of Western Australia, which he speculates was induced by a bolide impact. As evidence, he cites a rock painting on a rock, called Comet Rock, near Kalumburu, which depicts a comet-like motif and lies 5 km from the ocean on a plain covered in a layer of beach sand. Aboriginal oral traditions provide descriptions of a deluge around Walcott Inlet (e.g. Mowaljarlai and Malnic 1993:180-181), covering mesas up to 500 m high (Bryant et al 2007). One such account (Lucich 1969:52-57) describes a Jauidjabaija woman leaving Montgomery Island (Yawajaba Island) and canoeing to an island near the "fountain of the sea" (presumably near the mouth of Walcott Inlet[f], as it lies ~25 km south of Yawajaba Island). The story tells that the woman was on the coast of the island when the ocean rapidly receded, leaving many sea animals beached. A rushing noise followed some time later and the ocean tide came rushing back, covering the land, including mountains. The flood lasted a day before returning to the sea. Bryant et al (2007) argues that such mega-tsunamis are probably caused by a cosmic impact in the Indian Ocean. Tides in the Kimberleys can vary by as much as 10 meters, but such a substantial flood, preceded by a rapid recession of the ocean, indicates a genuine tsunami and not merely a higher-than-normal tide.

An Aboriginal elder from the Kimberleys tells a how when a man dies, his spirit is carried to Meteor Island[f] where he sits upon a long, chimney-like rock, regarded as a meteorite (Mowaljarlai and Malnic 1993:159-161). He turns away from his former life, placing his foot on the rock, causing it to overturn and crash into the sea shaking the earth, while his family chants "*Kaaaa-o*". He then shoots like a rocket into the spirit world. Mowaljarlai and Malnic claim that Meteor Island[f] and Entrance Island[f] are so prone to meteorite falls, that you can simply go out and collect meteorites on the ground and in the sea with minimal effort. No coordinates were given for Meteor Island, but text states that it "lies near Augustus Island[f] (~90 km northeast of Yawajaba Island) further out from Port George[f]" (*ibid*:159).

A Dreaming story of the Karajaree in the Kimberleys tells how the culture hero Miriny,

while wandering with his wives, Wade and Gololo, heard the noise of a bullroarer (*bolewana*) in the sky and saw tjuringas (*gaellgoro* - sacred stones) falling down from heaven (Worms 1943/44:297)[7], which may describe a meteorite fall (Event #17). The Worora, just north of Derby, have a children's story about the moon falling to earth onto two children as punishment for staring at the moon (it was taboo for children to stare at the moon). At the site are two stone pillars representing the children (Lucich 1969:33-34).

### 3.4    Tasmania

A story from Bruny Island in southeast Tasmania tells how two star-spirits fought each other and fell to earth in Louisa Bay[f], where one turned into a large stone that can still be seen today (Coon 1972:288, Event #18). A similar account from the nearby Needwonee is given in Haynes (2000:57). A story from Oyster Bay[f] tells how fire was brought to the earth by two men who stood on a mountaintop and "threw fire, like a star" that "fell among the blackmen" (Robinson 1966:95). The two men live in the clouds and can be seen in the night sky as the stars Castor and Pollux (the Gemini twins). The concept of fire being brought to earth by sky spirits in the form of a falling star is also found among the Yolngu of Arnhem Land, as described previously. Although a direct comparison is made between fire and stars, the story states that fire was thrown *like* a star, not *as* a star.

### 3.5    Queensland and Victoria

While Queensland and Victoria are both rich in stories and descriptions of comets and meteors (see Hamacher and Norris 2010a, 2010b), there were no Aboriginal stories or descriptions of meteorite falls or cosmic impacts in the reviewed literature for either state, except for brief colonist accounts of the Tenham meteorite in Queensland and the Cranbourne meteorites in Victoria, as described below. To date, only two impact craters are known in Queensland and both are considerably large and old (see Table 1a). Victoria has no confirmed impact craters.

However, there is an account from the Torres Strait Islands, which are administered by Queensland, of a stone that fell from the heavens (Event #19). The story tells of people camping on the tiny islet of Pulu[8] (near Mabuiag Island[f]). A "great stone" called *Menguzi kula* fell from the sky, crushing everyone except two lovers, who then became the progenitors of the current population (see Haddon 1904:22; Róheim 1971:370-371; McNiven et al 2009:311-313). Torres Strait Islanders are distinctly different from Aboriginal Australians, as they are of Melanesian extraction, having a close cultural and genetic link with Papuans.

### 4.0    Meteorites Falls and Finds

A meteorite fall may be a source of fear or fascination for the peoples who witness the event. Poirier (2005:237-238) cites an example of a meteorite fall near Jupiter Well[f], Western Australia that was incorporated into a new storyline, while Madigan and Alderman (1939:355-356) tell how Aboriginal people would steer clear of the Huckitta meteorite[9] and suggested that the they were in awe of the stone, perhaps considering it sacred. In 1879, a shower of meteorites fell near Tenham[f] station in South Gregory, western Queensland (Hellyer 1971). It was reported[10] in 1900 that Aboriginal people were "deadly afraid" of the Warbreccan masses of the Tenham meteorite: "They cover them in the bush with kangaroo grass, a twisted gidga bark and mud, and then by boughs over the top. Their idea is if the sun sees them, more stones will be shaken down to kill them", suggesting that they had witnessed the fall (Event #20).

---

[7] We were unable to obtain the original text from Worms (1940).
[8] Pulu is a tiny islet, with an area of $< 0.5$ km$^2$ and is considered a sacred place (see McNiven et al 2009).

According to Spencer (1937) the Warbreccan stones were taken by an opal dealer named T.C. Wollaston and sold to the British Museum, using an invented story to explain their origins and how he acquired them. The Aboriginal account is taken from files in the British Museum, so this account is considered dubious (although we do not know if the account is fictitious). However, a news article (Anon 1880) describes an eyewitness account of the Tenham fall on Monday, 25 April 1880 that involved an Aboriginal policemen and their reaction to the event:

> "A few minutes after six o'clock […] a very large and brilliant meteor shot from overhead and descended in a southerly direction. The meteor appeared to be the size of a six-quart billycan, and was one splendid ball of fire; it left no streak of light after it, and was the largest one I have ever seen. When the meteor had descended about three parts the distance from where I first saw it to the earth, I lost sight of it, as it was passing behind a large dark cloud. I stood looking in the direction the meteor was traveling, when a loud explosion took place in the same direction which slightly shook the ground for miles around; then a loud rushing noise could be heard as though a great blast of air was rushing through a large tube suspended in mid-air. This sound must have lasted for nearly two minutes when it died away. Next morning, when I rode up to Jundah, everyone there wanted to know what the explosion was, and the only conclusion we could arrive at was that when the meteor struck the ground it must have exploded, but we have not been able to account for the rushing sound afterwards. Inspector Sharp of the black troopers [an Aboriginal police force] said that when the explosion took place, the house he was in shook very much, and that when he ran out to see what was taking place he saw all the troopers running into the barracks with fright depicted on each countenance. From what I could learn I was the only white man at Jundah who saw the transit of the meteor. It is my opinion that it struck the earth a few miles above Galway Downs, and close to the Barcoo River, or we could not have felt the earth shake when it exploded."

There were conflicting accounts and dates of the fall recorded in the media (see Spencer

---

[9] The Huckitta meteorite was discovered by an Aboriginal worker named Mick Laughton.
[10] British Museum of Natural History in the meteorite files under "Jhung" and "Tenham".

1937). Although the account states that the Aboriginal policemen were struck with fright, this is an expected reaction for anyone who witnesses such an event.

The Yintjingga people of Stewart River in the Kimberleys perceive a meteor as a spirit (*mipi*) and a portent that someone has died. Sometimes the mipi "brings his light" and crashes to the earth, creating noise. At the same moment, a "big devil" (Wo'odi Mükkän) that sits in a tree drops a "great stone" to the ground (Thompson 1933:499). This provides direct links between the meteor, the noise made upon impact, and the presence of a stone at that spot, presumably a meteorite. Thompson notes that the Aboriginal ideas of the stone being dropped from the sky "can be easily understood, for meteorites must at times have been seen actually to fall and bury themselves in the earth" (*ibid*:499). The Mycoolon of northwest Queensland believe that death resulting from a stone falling from the clouds is a penalty for children eating forbidden food (Palmer 1884:294).

However, not all meteorites were viewed negatively. The Mycoolon believe that falling stars strike and penetrate certain Acacia trees, transforming into gum (sap) that is a well-liked food source (*ibid*). Bevan and de Laeter (2002:18) mention how a group of Aboriginal people would dance around a larger fragment of the Cranbourne[f] meteorite fall (southeast of Melbourne) while hitting it with stone axes, apparently "pleased by the metallic sound it made". The meteorite may have been part of a ritual or ceremony of some kind, which was either unknown or unacknowledged by the colonists. It may also be that the Aboriginal people interacting with the Cranbourne meteorite did not see it fall and had no reason to fear it, as opposed to the Jupiter Well and Tenham meteorites. Meteorites have been found in deserted Aboriginal camps (e.g. Johnson and McColl 1967; Alderman 1936:542), but there is little hard evidence to determine whether the stones had any special significance or practical use.

### 4.1    Aboriginal Discovery of Meteorites and Impact Craters

Over the last century, Aboriginal people made significant contributions to the discovery of meteorites and impact craters in Australia. A lunar meteorite was found by an Aboriginal meteorite hunter in the Millbillillie[f] strewnfield in the Nullarbor near Calcalong Creek in Western Australia (Wlotzka 1991; Hill et al 1991). According to Hill et al (1991:614), the creek derives its name from an Aboriginal word meaning "seven sisters (Pleiades) went up into the sky, chased by the Moon", although this is not sufficient evidence that the meteorite has anything to do with the name of the creek. Dalgaranga Crater, located 75 km west of Mount Magnet in Western Australia, was discovered in 1923 by an Aboriginal man named G.E.P. Wellard[11] (McCall and de Laeter 1965:28,60) but was not identified as an impact crater until 1938 (Simpson 1938). An Aboriginal man named Billy Austin witnessed a bright meteor around 05 February 1924 and later discovered a crater in the central South Australian desert near Lake Labyrinth[f] approximately four meters in diameter, with the impacting meteorite 33 meters to the west (Spencer 1937:353-354). Numerous additional meteorite discoveries have been made by

---

[11] Wellard was also instrumental in the identification of other meteorites, including the 33.6 kg Gnowangerup meteorite (de Later 1982), a Mellenbye/Yalgoo meteorite fragment (McCall and de Laeter 1965:38).

Aboriginal people and are well recorded in the literature (see Hodge-Smith 1939; Madigan and Alderman 1939:353; Barker 1964:109-110; McCall and de Laeter 1965:28,33,38,39,41,52).

## 4.2    Aboriginal Use of Meteoritic Material

Bevan and Bindon (1996) published a comprehensive work on the Aboriginal use of meteoritic material, showing that there was no evidence that Aboriginal people worked meteoritic iron, despite some accounts of Aboriginal people using meteoritic iron for weapons (e.g. Peck 1925:152).  There is evidence to support the transport of meteoritic material by Aboriginal peoples (*ibid*), but little physical evidence to explain the use of meteorites or their significance to Aboriginal cultures.

However, the use and role of tektites (glass-like terrestrial rocks formed by meteor impacts) among Aboriginal peoples was studied extensively by Baker (1957) and Edwards (1966), who showed many Aboriginal groups had a number of uses for tektites, including surgical tools and implements used in ritual and ceremony.  Some Aboriginal groups in Western Australia and the Nullarbor Plains believed tektites to be magic "sky stones" and associated them with meteors and cosmic impacts (Bates 1924:11d).

Gibbons (1977) presents a comprehensive (albeit dated) list of meteorites, impact structures, and meteorite fall/find locations (as of the late 1970s) which can be used, in conjunction with current databases, to correlate associated Dreamings with known meteorite finds or cosmic impacts.

**Table 2:**  Coordinates of the sites of cosmic impact stories or meteorite falls/finds described in this paper, listed alphabetically, shown in Figure 3.

| Name | Latitude | Longitude | State |
|------|----------|-----------|-------|
| Augustus Island | 15° 20′ S | 124° 34′ E | WA |
| Bickerton Island | 13° 46′ S | 136° 12′ E | NT |
| Bodena | 29° 25′ S | 146° 40′ E | NSW |
| Caledon Bay | 12° 46′ S | 136° 28′ E | NT |
| Cranbourne | 38° 06′ S | 145° 16′ E | VIC |
| Entrance Island | 16° 16′ S | 124° 37′ E | WA |
| Eucla | 31° 42′ S | 128° 50′ E | WA |
| Forrest | 30° 51′ S | 128° 06′ E | WA |
| Girilambone | 31° 14′ S | 146° 53′ E | NSW |
| Hermannsburg | 23° 56′ S | 132° 46′ E | NT |
| Huckitta | 22° 22′ S | 135° 46′ E | NT |
| Jupiter Well | 22° 52′ S | 126° 35′ E | WA |
| Kurra Kurran | 33° 02′ S | 151° 38′ E | NSW |
| Lake Labyrinth | 30° 20′ S | 134° 45′ E | SA |
| Lake Mackay | 22° 35′ S | 128° 41′ E | WA |
| Louisa Bay | 43° 29′ S | 146° 14′ E | TAS |
| Mabuiag Island | 09° 56′ S | 142° 10′ E | QLD |
| Marabibi | 12° 37′ S | 131° 40′ E | NT |
| McGrath Flat | 35° 52′ S | 139° 24′ E | SA |
| Meteor Island [a] | 15° 19′ S | 124° 37′ E | WA |
| Millbillillie | 26° 27′ S | 120° 22′ E | WA |
| Mount Doreen | 22° 06′ S | 131° 26′ E | NT |
| Oyster Bay | 42° 09′ S | 148° 06′ E | TAS |

| Port George | 15° 22′ S | 124° 42′ E | WA |
| Puka (Palm Valley) | 24° 03′ S | 132° 43′ E | NT |
| Tenham | 25° 44′ S | 142° 57′ E | QLD |
| Wilcannia | 31° 33′ S | 143° 30′ E | NSW |
| Walcott Inlet | 16° 25′ S | 124° 34′ E | WA |
| Yurntumu | 22° 15′ S | 131° 47′ E | NT |

ᵃ Coordinates are of the area between Entrance and Augustus Islands.

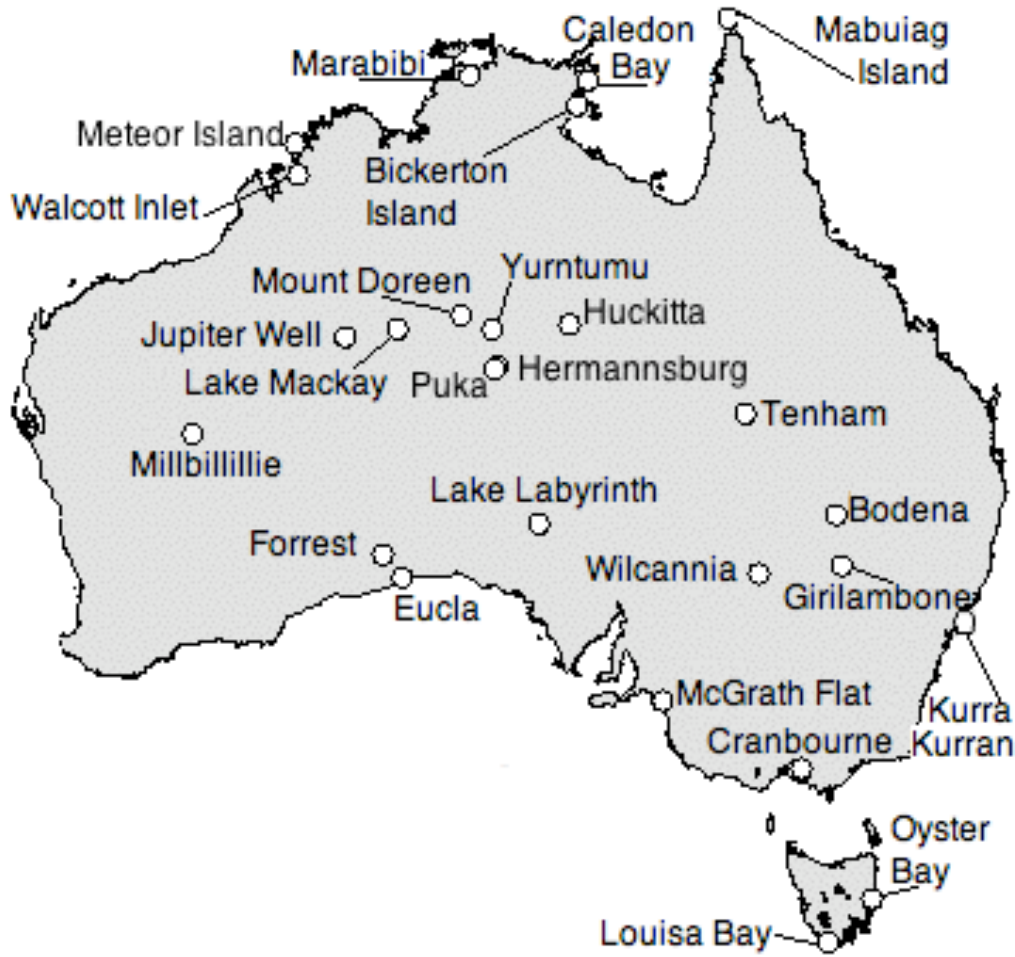

**Figure 3:** Sites of cosmic impact stories or meteorite falls/finds described in this paper. The coordinates of these sites are given in Table 2.

## 5.0    Statistics of Meteorite Falls and Cosmic Impacts

We now attempt to quantify the meteorite influx rate in Australia, the probability of a person seeing a meteorite-producing meteor from any location in Australia, and the probability of a person being hit or killed by a meteorite or airburst (exploding meteor).

### 5.1    Meteorite Influx Rate

It is difficult to quantify the rate of meteorite falls on earth's surface given the number of variables and uncertain statistics. However, general estimates can be made using different techniques. One technique is to observe the number of bright meteors over a given area combined with the meteorite recovery rate (see Halliday et al 1984). The Halliday et al technique derives an annual fall rate, per million km$^2$, of 39 for a minimum mass $m > 0.1$ kg, 7.9 for $m > 1$ kg, and 1.6 for $m > 10$ kg. Another technique is to derive a fall rate based on recovered meteorites within a surveyed area (see Zolensky et al 1990). Using a sample of recovered meteorites within an 11 km$^2$ area in New Mexico, Zolensky et al provides a fall rate of ~940 meteorites per million km$^2$ per year with a mass exceeding 10 g. Table 3 lists the expected meteorite fall rates over Australia and each of its states and territories, given these estimates.

**Table 3:** The expected meteorite fall rate based on estimates given by Halliday et al (1984) and Zolensky et al (1990). Land areas are given in $10^6$ km$^2$, sorted by minimum mass (kg). $N_{yr}$ is the fall rate per year. State abbreviations are the same as in Table 1, while Australia is abbreviated as AU. Fall values are rounded to the nearest whole number.

| State | Area | $N_{yr}$ per minimum mass (kg) | | | |
|-------|------|------|------|------|------|
|       |      | 0.01 | 0.1 | 1.0 | 10.0 |
|       | $10^6$ | 940 | 39 | 7.9 | 1.6 |
| AU  | 7.617 | 7144 | 296 | 60 | 12 |
| WA  | 2.529 | 2378 | 99 | 20 | 4 |
| QLD | 1.730 | 1626 | 67 | 14 | 3 |
| NT  | 1.349 | 1268 | 53 | 11 | 2 |
| SA  | 0.983 | 921 | 38 | 8 | 2 |
| NSW | 0.800 | 752 | 31 | 6 | 1 |
| VIC | 0.227 | 213 | 9 | 2 | 0 |
| TAS | 0.068 | 64 | 3 | 1 | 0 |

During known human habitation of Australia (a lower limit of 40,000 years), an estimated 286 million meteorites with mass > 10 g have fallen. During periods of lower sea levels, this number is higher because of the larger continent size. While Aboriginal language groups are not spread evenly in size and population density across the continent, a rough approximation of 300 distinct Aboriginal language groups (Walsh 1991) of equal size gives a land area of ~25,300 km$^2$ per group, showing that each group would expect approximately 23 meteorite falls per year within their land area, each assumed to be accompanied by a visible meteor[12].

## 5.2 Probability of Casualty due to Meteorite Falls and Cosmic Impacts

### 5.2.1 Probability of Being Struck by a Meteorite

Historic accounts of meteorites striking or killing people are recorded in the literature (e.g. Yau et al 1994; Gritzner 1997; Lewis 1999:14-25). Accounts from Aboriginal sources include Mountford (1976:457) who noted that men in Western Australia had been killed by fiery stones thrown to the earth by the Walanari. However, meteorite falls are a relatively rare occurrence at any given location, so what is the probability that a person has been struck by a meteorite since humans first arrived in Australia?

---

[12] However not all visible meteors produce recoverable meteorites, as many burn up in the atmosphere.

Given the number of meteorites that fall per annum as described in the previous section over the land area of Australia ($N_m$ = 7144 for m > 10g, from Table 3) and assuming an evenly distributed constant population of 250,000 people ($N_p$) over the course of human history in Australia ($T_A$ = 40,000 years), we estimate the number of people who have been struck by a meteorite in Australia ($N_s$) as:

$$N_s = \left( \frac{A_H}{A_A} \right) N_m \ N_p \ T_A$$

(1)

Where $A_H$ is the surface area of a person (from a "bird's eye view", estimated at 0.3 m$^2$) and $A_A$ is the area of Australia (7.6x10$^{12}$ m$^2$). For the values given, the number of human beings struck during the course of Australian human history is about three. Because this happens so infrequently, it is extremely unlikely that the literature would have records of such an event in Australian human history, yet there are several Aboriginal accounts of people being killed by meteorites. Of course, populations are not spread evenly across the continent and populations fluctuate in number, so this and all subsequent estimates in this paper serve only as first-order approximations.

### 5.2.2    Probability of Death or Injury from Cosmic Impacts

To estimate the probability of being within the destruction area of an airburst or impact in Australia, we first determine the area $A_D$ (km$^2$) of devastation caused by an impactor of energy $E$ (Mt) using the relationship given by Steel (1995):

$$A_D(E) = 400 \ E^{0.67}$$

(2)

Thus an impact energy equivalent to the Hiroshima bomb (13 kt) would destroy an area of ~22 km$^2$. Impactors with energies of 1 Mt will devastate an area of 400 km$^2$, equivalent to the area of a large city. To estimate the frequency of impacts by impact energy, we estimate the time interval between impacts $t$ (years) of a given impact energy $E$ (Mt), using Collins et al (2005: their Equation 3):

$$t_i(E) = 109 \ E^{0.78}$$

(3)

The 1908 Tunguska airburst in Siberia was estimated to have an energy of 10 Mt from a height of 8.5 km and destroyed ~2000 km$^2$ of Siberian forest, in agreement with Equation 2 (Napier and Asher 2009). Using Equation 3, Tunguska-like events are expected about every 650 years. However, other large bolide explosions have occurred within the last century, including the burst over the Curuçá River, Brazil in 1930 (Bailey et al 1995; Huyghe 1996) with an estimated energy of 1 Mt (Chown 1995) and a Tunguska-like blast over Guyana in 1935 (Steel 1996). None of these airbursts left a known impact crater or meteorite debris, which is consistent with a cometary origin (since comets are composed of low-density ice and dust). Additionally, all three airbursts occurred when the earth passed through major meteor streams (Napier and Asher 2009), caused by the earth passing through the dust tail of a comet.

However, the atmospheric dynamics of meteors are still poorly understood and recent research indicates that the Tunguska blast had much lower impact energy than previously estimated (3-5 Mt, see Boslough and Crawford 2008). Estimates of the time interval of Tunguska-like impacts range from once every 1,000 years (Brown et al 2002) to once every 250 years (Equation 3). Networks scanning the global atmosphere for explosions were not incorporated until the latter 20th century, so other airbursts could have occurred over the ocean unnoticed. Rocky meteoroids, corresponding to impact energies of up to 1 Mt, often explode high in the atmosphere, having little effect on the earth (Collins et al 2005). Figure 4 shows the estimates of the impact frequency and impact energy of impactors based on several Near Earth Object (NEO) surveys. The assumptions and equations used in this paper provide a reasonable first-order approximation for estimating impact probabilities.

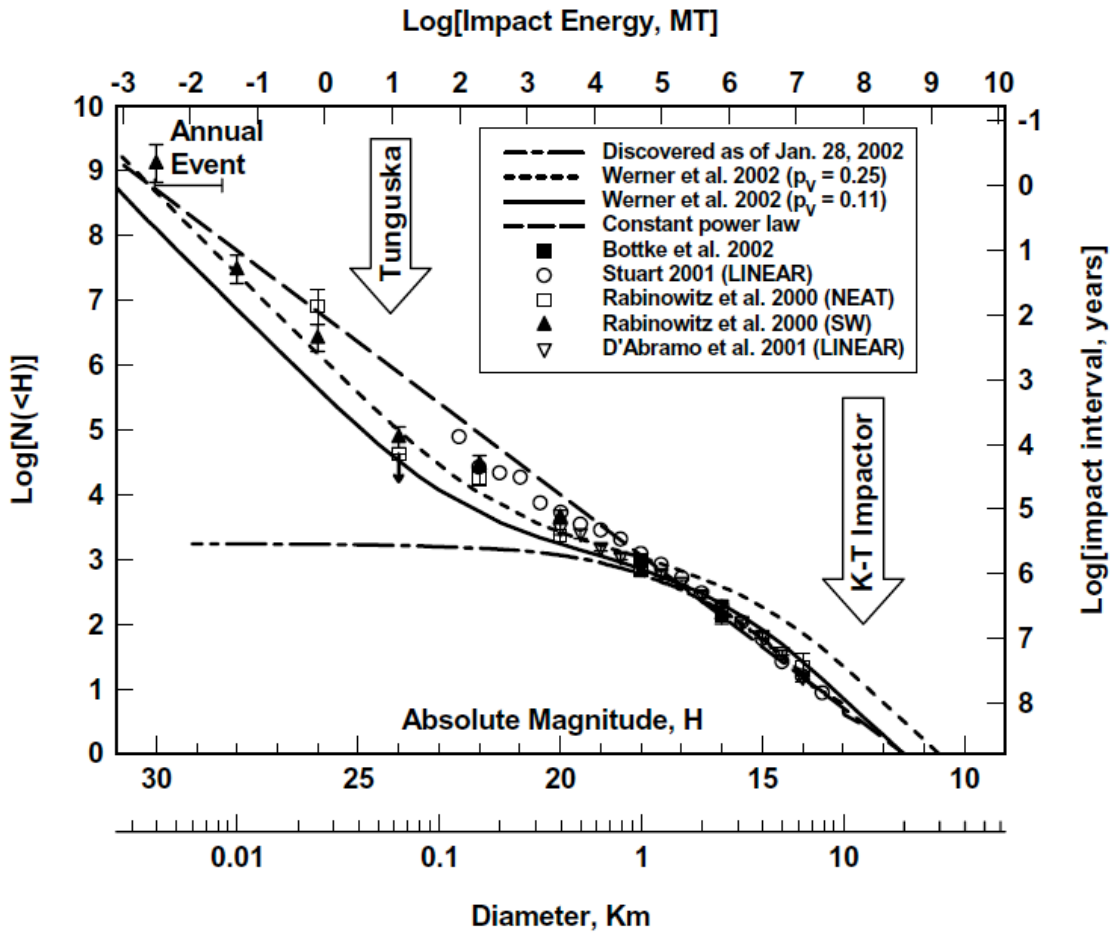

**Figure 4:** The impact frequency and energy of Near Earth Objects from several surveys (taken from Morrison et al 2003, with corrected publication dates). The horizontal axis shows the impact energy, and also the meteoroid diameter and absolute magnitude (luminosity) of the NEO. The vertical axis shows the time interval between such impacts and also the expected number of NEOs that are brighter than a given magnitude (H). A review of this subject can be found in Ivanov (2008).

The time between impacts within Australia, using a modified version of Equation 3, is given as:

$$t_{i_A}(E) = 109 \, E^{0.78} \left( \frac{A_E}{A_A} \right)$$

(4)

Where $A_E$ is the area of earth ($5.1 \times 10^8$ km$^2$) and $A_A$ is the area of Australia ($7.6 \times 10^6$ km$^2$). Equation 4 tells us that a 1 Mt event occurs about every 7,300 years. This means that approximately 5.5 such events have occurred in Australia over the last 40,000 years.

The total number of people killed by any single 1 Mt event in Australia is:

$$N_{K_i} = \left( \frac{A_D(E)}{A_A} \right) N_P$$

(5)

Where $A_D(E) = 400$ km$^2$ (from Equation 2), $A_A$ is the area of Australia and $N_P$ is the total average population of Australia (250,000). Equation 5 shows that ~13 people are killed by any single 1 Mt event.

Since ~5.5 such events have occurred during the human habitation of Australia, each killing 13 people, the total number of people killed by all combined impact events is about ~72.

Of course, such events would have had an influence (perhaps distress or surprise) on the people in the region that witnessed the event but were not killed. For a 1 Mt event, the radius of destruction (Equation 2), is ~11.3 km. If we assume that people within a 25 km radius heard or saw the event, then the "area of influence" would be ~2,000 km$^2$. This would mean only about 360 people have been "influenced" by such an event during the course of human history in Australia. However, this number should be taken as no more than a rough estimate, since it will depend on how the population is distributed.

Another approach is to estimate the number of language groups that would have witnessed an impact. If we assume 300 evenly spread language groups and estimate that five impact events (with E = 1 Mt, using Equation 4) have occurred during the human history of Australia, then only five language groups would have witnessed an impact over the last 40,000 years. Over the last 10,000 years, that number drops to only one. If oral traditions can survive for 10,000 years, we would statistically expect to find only one story. Since many Aboriginal cultures, including their oral traditions, were damaged or lost because of colonisation, it would be surprising if any group now had an oral tradition based on a witnessed cosmic impact.

## 6.0   Discussion

We now compare the impact rates predicted by the meteor literature with those indicated by Australian Aboriginal tradition. The results are summarized in Table 4. We emphasise that extreme caution is needed when evaluating such subjective material quantitatively, and that the following discussion should be treated as a guide to further research rather than as a definitive statement.

Table 4 shows that the numbers of events expected from meteoritic knowledge is comparable to that recorded in oral tradition, with the exception of witnessed destructive events, where the number of stories is about three times higher than that expected. However, the oral traditions sampled here are only a small fraction of those that have originated over the 40,000 years of Australian human history. While the longevity of oral traditions is not known (and remains the topic of debate), we have examples that suggest these records can last thousands of years, including those of the erupting crater lakes in Queensland (Dixon 1972:29), the eruption of Mount Gambier in South Australia (Anon 1870), and the traditions associated with the Henbury impacts (P&WCNT 2002:15), (see Blong (1982) for examples from Papua New Guinea, and Masse and Masse (2007) for examples from South America). If we liberally estimate that Dreaming stories have a typical longevity of 10,000 years and that perhaps a quarter of pre-contact Aboriginal cultures have had their oral traditions recorded and published (many stories are considered sacred and secret and are not shared with outsiders), then the number of stories describing cosmic impacts exceeds the number expected by a factor of about 17, or, in the case of witnessed major impacts, by a factor of about 50.

The explanation of this discrepancy may include a combination of the following factors:

      (1) These stories were not based on witnessed events;
      (2) These stories have been influenced by Western science;
      (3) A single story was shared with many other Aboriginal groups;
      (4) Impact events were more frequent than estimated;
      (5) Some unacknowledged factor plays a role.

In the following sections, we address each topic separately.

**Table 4:** Comparison of numbers of meteor events predicted by the meteor literature throughout human habitation of Australia (last 40,000 years), compared with those indicated by Australian Aboriginal tradition.

| Phenomenon | Meteoritic/Geological evidence | Oral tradition | Event # | Notes |
|---|---|---|---|---|
| Known impact craters in Australia formed within span of human habitation. | Three craters known (Boxhole, Henbury and Veevers). | No oral tradition of being caused by a cosmic impact, except perhaps Henbury where the stories are secret. | | Two other craters (Gosse's Bluff and Wolfe Creek) have impact stories but were formed before human habitation. |
| Meteorite strikes and kills a human | ~3 fatalities expected | 1 or 2 fatalities in oral tradition | 16?, 19 | |
| Meteorite impacts causing human fatality | About 5 events expected, | 7 oral accounts of fatal events | 1, 3, 7, 8, 12, 16, 19 | |
| Meteorite impacts causing destruction or crater, but not necessarily human fatality | Expect about 5 language groups to have witnessed such an event | 17 events (including those involving fatalities) are recorded in the oral tradition. | Those above, plus 4, 5, 6, 9, 10, 11, 13, 15 | |
| Notable bright meteorite impacts but not causing destruction | Many expected – difficult to quantify. | | 2, 14, 17, 18, 20 | |

## 6.1 Stories were not based on witnessed events

In some cases, the description is vague (see Bevan and Bindon 1996:96; Lucich 1969:33-34), but in other cases, the description closely parallels the scientific explanation (see Peck 1933:192-193; Harney and Elkin 1949:29-31). The account of a cosmic impact near Wilcannia described by Jones (1989) in Section 3.1 is an example of a story that claims the event was witnessed but currently lacks physical evidence to support it (see Steel and Snow 1992:572). It is possible that the event occurred elsewhere, at a location yet to be identified by Westerners (Bevan and Bindon 1996:95). While this explanation may be the case for some stories, other descriptions of impact events so closely parallel the scientific explanation (e.g. Peck 1925, 1933) that it seems unlikely that the story has been fabricated. However, because a Peck, a Westerner, recorded the stories, it is unknown how much his bias affected his writings.

It is also possible that the impact stories derived from logical deductive reasoning. For example, dropping a rock into the sand may have helped people recognize that larger falling rocks may have created larger craters that are of a similar shape. A smaller impact, such as the one witnessed at Lake Labyrinth (Section 4.1), could have prompted elders to create a story about a similar event.

## 6.2 Stories have been influenced by Western science

Another possibility is that these accounts were influenced by Western science, with Western information about cosmic impact events being incorporated into oral tradition.

This is certainly possible in some cases. For example, the literature records no accounts of the Wolfe Creek crater having cosmic origins before it was known as an impact crater (see Sanday 2007 versus Bevan and McNamara 1993:6). However, many of the Aboriginal informants interviewed by Sanday (2007) and Goldsmith (2000) maintain that their accounts of the cosmic origins of Wolfe Creek (Kandimalal) were handed down over many generations, before Westerners knew of its cosmic origins. This would be difficult to substantiate unless a written record of the story existed before the crater was identified by Westerners. Unfortunately, in the case of Gosse's Bluff and Wolfe Creek, no written records have been found describing these structures prior to their identification as impact craters by Western science.

### 6.3    A single story was shared with other Aboriginal communities

Examples of a particular story that was shared by several distinct communities exist, of which some were described in Section 3.1, including the Mathews (1994) account and the McKay (2001:112-114) account. In this case, a single group witnessed an event and the story of that event spread to other communities. However, this suggests that *some* group must have witnessed the event. It is possible that many stories are based on a few much earlier stories, which have spread to other communities and placed in a local context (see also Michaels 1985). The oldest recorded story may identify the particular community from which it came.

### 6.4    Impact events were more frequent than estimated

If the Aboriginal accounts described in this paper are records of witnessed events, despite the results of Section 5, then the evidence suggests that we are underestimating the impact rate. This could either be because the literature on impact rates systematically underestimates rates, or because impact events are more frequent than previously thought.

The distribution of known impact structures is not uniform, For example, 25 craters are found in the Baltic countries and Scandinavia, while the rest of Western Europe contains only four. The remote, dry Australian continent contains 27 confirmed craters, while the much larger and wetter South American continent contains only eight. This indicates that many more impact craters remain to be discovered. In remote, moist areas such as the rainforests of the Amazon, Congo and Southeast Asia, no confirmed craters have been found. On the ocean floor, which should account for ~70% of impact craters, the search for these structures is difficult.

The hypothesis that impacts are more frequent than currently estimated was proposed by members of the Holocene Impact Working Group[13], which claims that some historic catastrophes were caused by impact events (see Baillie 2007; Masse et al 2007b; Firestone et al 2006; Rappenglück et al 2009[14]). There are only 11 confirmed Holocene (i.e. within the last 12,000 years) impact sites out of 176 known craters to date (Earth Impact Database 2009, see Table 4). Figure 4 shows the estimated impact intervals based on impact energy and impactor diameter. Using the scaling equations of Collins et al (2005), we expect large impact events (impactors exceeding 1 km diameter) to occur at intervals greater than 100,000 years.

---

[13] http://tsun.sscc.ru/hiwg/hiwg.htm

[14] A full list of publications by the group can be found here: http://tsun.sscc.ru/hiwg/publ.htm

However, proposed structures from the Australian-New Zealand region suggest multiple large impacts have occurred over the last few thousand years. These include the Mahuika structure described in Section 1.3, as well as two proposed submarine structures, named Kanmare (meaning 'Serpent') and Tabban (meaning 'Rainbow'), that were discovered either side of Mornington Island in the southeast corner of the Gulf of Carpentaria (see Figure 5, Table 5, and Appendix B). The latter two structures are approximately 18 km and 12 km in diameter, respectively, based on satellite altimetry and have a measured date (first order approximation) of 572 CE ± 86 years (Abbott et al 2007)[15]. Numeric models of these impacts (using Collins et al 2005) show that Mornington Island and surrounding archipelago would have succumbed to devastating effects (including widespread fires, impact ejecta, tsunamis and earthquakes), of which we should expect to find significant archaeological, geological, and environmental evidence. Another alleged submarine structure found in the southern Indian Ocean, named Burckle, has a diameter of ~ 30 km and a proposed age of < 6,000 years (Abbott et al 2009).

In addition, there are several other large, alleged impact structures from the other regions of the world that have proposed dates within the last 6,000 years (e.g. Umm al Binni structure; see Master and Woldai 2004, Chiemgau crater field; see Rappenglück et al 2009)[16]. To account for this discrepancy, Baillie (2007) suggests these impacts may have originated from fragmented comets or "dark" comets, which have a very low albedo and are difficult to detect (Napier et al. 2004) and that current impact estimates represent very conservative minima.

**Table 4:** Confirmed Holocene impacts (Earth Impact Database 2009), see also Baillie (2007). Provided are the names, locations, ages. Crater fields are noted as well as any impact sites with related myths or records of the event.

| Name | Country | Age (years) | Field | Associated Myths/Records |
|------|---------|-------------|-------|--------------------------|
| Sikhote Alin | Russia | 63 | No | Gallant (1996) |
| Wabar | Saudi Arabia | 140 | Yes | Bobrowsky and Rickman (2007:30) |
| Haviland | United States | < 1000 | No | - |
| Sobolev | Russia | < 1000 | No | - |
| Whitecourt | Canada | < 1100 | No | - |
| Camp de Cielo | Argentina | < 4000 | Yes | Bobrowsky and Rickman (2007:31) |
| Kaalijärv | Estonia | 4000±1000 | Yes | Bobrowsky and Rickman (2007:29) |

[15]Tester and Abbott (2007) describe the craters as elliptical in shape, indicative of a low angle impact (< 15°).

[16]Another close pair of submarine crater candidates consist of the 5 km wide Kangaroo and 4 km wide Joey structures (Abbott et al 2006), located south of Portland, Victoria (see Figure 5-Bottom, Table 5). Geophysical surveys of these structures are in progress (e.g. Martos et al 2006; Elkinton et al 2006; Abbott et al 2006), although no age estimates have been published. A news account (Anon 1870) of an Aboriginal story describing a massive tsunami that destroyed the region between Portland and Mt. Eckersley, Victoria suggests a potential link to the Kangaroo and Joey craters, but includes references to volcanic activity on Mt Gambier, suggesting either a volcanic event or two separate events incorporated into the same story. Any connection between the craters and the Aboriginal account is speculative.

| | | | | | |
|---|---|---|---|---|---|
| Henbury | Australia | 4200±1900 | Yes | Bobrowsky and Rickman (2007:31) | |
| Ilumetsä | Estonia | < 6600 | No | Bobrowsky and Rickman (2007:29) | |
| Macha | Russia | < 7000 | Yes | - | |
| Morasko | Poland | < 10000 | Yes | - | |

**Table 5:** Submarine crater candidates from the Australia-New Zealand region, shown in Figure 5. Provided are the location, estimated diameter (D), and proposed age (T) of each structure.

| Name | Latitude | Longitude | D (km) | T (years) | Location |
|---|---|---|---|---|---|
| Kanmare | 16° 34′ S | 139° 03′ E | 18 | 572 CE ± 86 | Gulf of Carpentaria |
| Tabban | 17° 07′ S | 139° 51′ E | 12 | 572 CE ± 86 | Gulf of Carpentaria |
| Kangaroo | 39° 02′ S | 141° 16′ E | 5 | Unknown | Off Coast of Victoria |
| Joey | 39° 09′ S | 141° 11′ E | 4 | Unknown | Off Coast of Victoria |
| Mahuika | 48° 18′ S | 166° 24′ E | 20 | ~1443 CE | Tasman Sea, south of NZ |
| Burckle* | 30° 51′ S | 61° 21′ E | 30 | < 6000 | Southern Indian Ocean |

\* Not shown in Figure 5

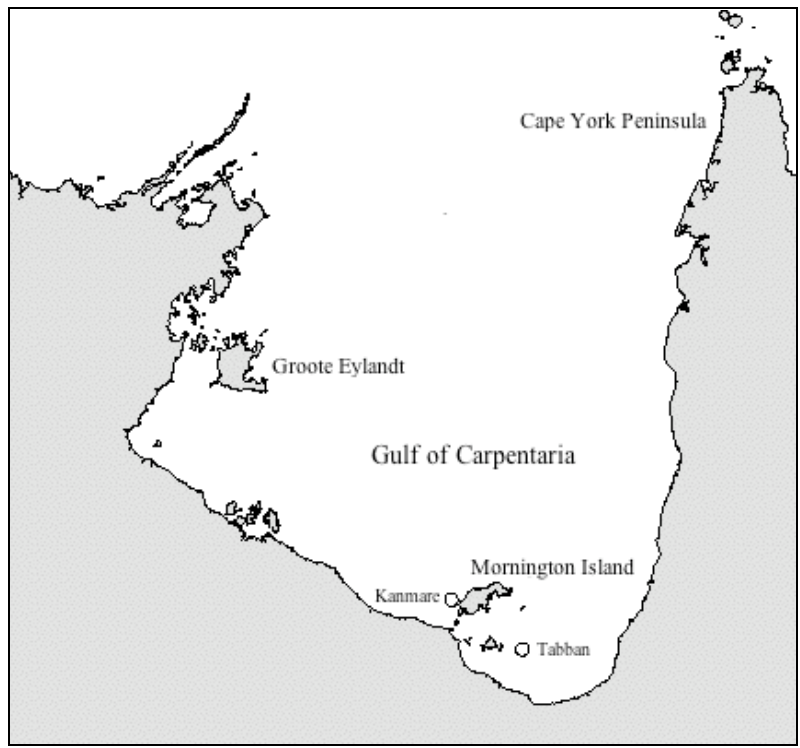

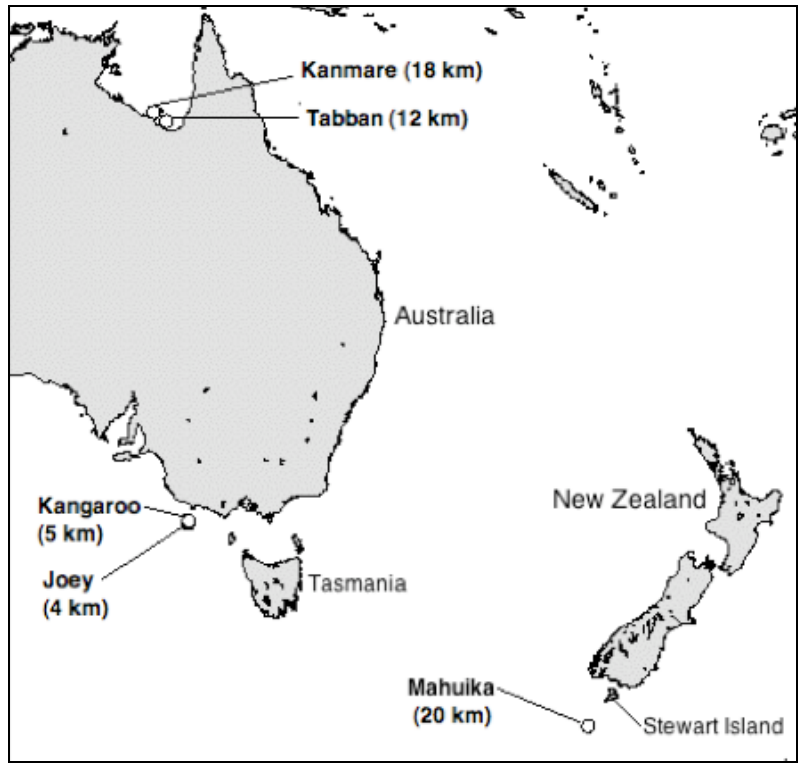

**Figure 5: Top** - The Kanmare and Tabban submarine crater candidates, located on either side of Mornington Island in the southern section of the Gulf of Carpentaria. **Bottom** - Proposed Holocene submarine impact craters near Australia and New Zealand, including the estimated crater diameter.

This hypothesis has been met with criticism and controversy and continues to be heavily debated (e.g. Baillie 2007; Gusiakov et al 2008; Pinter and Ishman 2008; Firestone and West 2008; Bunch et al 2008; Bourgeois and Weiss 2009; Rappenglück et al 2009; Abbott et al 2010).

## 7.0    Conclusion

None of the events described in Aboriginal oral traditions correspond to currently known impact events that occurred during the human history of Australia. Furthermore, although some impact events caused craters during Australian human history (e.g. Boxhole, Veevers, Henbury), there is no (publicly available) oral tradition of witnesses to these events.

On the other hand, we find that the number of stories describing eyewitness impact events in the oral traditions greatly exceeds the number expected from current estimates of meteorite influx rates and impact intervals. We conclude that either the current meteor literature severely underestimates the impact rate (and there are hotly contested arguments supporting this) or else the stories are not based on witnessed events.

We examine three possible explanations for why the stories may not be based on witnessed events: 1) the stories are post-contact and have incorporated information from Western science, 2) that the Aboriginal people independently deduced that impact craters were formed from a cosmic impact, or 3) that a single or small number of events were incorporated into oral traditions and spread to other communities across the continent. While we cite evidence supporting each explanation, we are unable to conclude with any certainty which explanation best explains the discrepancy.

We suggest that oral traditions may be used to identify impact structures and meteorites unknown to Western science. Besides the plethora of useful scientific information that would benefit various scientific fields, the information gathered from an impact structure described in Aboriginal oral tradition, including its age and impact effects, could show how some oral traditions evolve over long periods of time. Such studies may also be useful in modeling alleged submarine Holocene impact events, such as Mahuika, Tabban, or Kanmare. For this reason, we encourage investigations and surveys of the areas where cosmic impact stories were recorded, which may shed light on this elusive problem.

### Acknowledgments

The authors would like to thank and acknowledge the Aboriginal elders and custodians of Australia, the Darug people (the Traditional Custodians of the land on which Macquarie University is situated), and the Australian Institute for Aboriginal and Torres Strait Islander Studies in Canberra. We would also like to thank Dallas Abbott, Tui Britton, Edward Bryant, Andrew Buchel, John Clegg, Paul Curnow, Kristina Everett, John Goldsmith, Steve Hutcheon, Richard Kimber, John McGovern, and Craig O'Neill. Hamacher was funded by the Macquarie University Research Excellence Scholarship within the Department of Indigenous Studies at Macquarie University. Maps in this paper were generated using the GMT (Generic Mapping Tools) software package developed by Paul Wessel and Walter H. F. Smith (see http://gmt.soest.hawaii.edu/).

### Appendix

**Table A:** Aboriginal groups discussed in this paper, including the group name (language, clan, dialect, or community name), the general area of that group, including the state(s). Groups noted with an (*) are not identified in the text by group name, but rather by geographic location or identifying landmark, such as a river or town. In these cases, the group given is taken from the Australian Institute for Aboriginal and Torres Strait Islander Studies map of Aboriginal Languages.

| Group | General Region | State |
|---|---|---|
| Alyawarre | Southeast Region | NT |
| Amangu* | Region 200-350 km NW of Perth | WA |
| Anula [17] | Carpentaria Gulf Region | NT |
| Andedja | Kimberleys | WA |
| Anguthimri* | Near Mappon, far NW of Cape York | QLD |
| Arrernte [18] | Central Australia, area of Alice Springs | NT |
| Awabakal* | Lake Macquarie/Newcastle Region | NSW |
| Badaya | Region east of Darwin | NT |
| Barkinji | Northwest Region | NSW |
| Binbinga | Borroloola area, near Gulf of Carpentaria | NT |

[17] aka Yanyuwa

[18] aka Arunndata, Aranda, Arunta

| | | |
|---|---|---|
| Bindal* | Townsville, Cleveland Bay | QLD |
| Birria | South-Western Region | QLD |
| Bogan | North-central region | NSW |
| Boorong | North-Western Region | VIC |
| Boonwurrung* | Area South-East of Melbourne | VIC |
| Bundjalung* | Richmond-Tweed Region | NSW |
| Darkinung* | Cessnock-Wollombi Region | NSW |
| Dhauwurd[19] | Portland/southwest region | VIC |
| Dieri* | North East Region | SA |
| Djaru (Jaru) | North East Region, SE of Kimberleys | WA |
| Djirbalngan* | Tully River Region | QLD |
| Euahlayi | North-Central Region | NSW |
| Giya* | Proserpine Area | QLD |
| Goreng* | Near Bremer Bay | WA |
| Gundidjmara | Southwest Region | VIC |
| Gunwinggu | Arnhem Land | NT |
| Gurudara | Region east of Darwin | NT |
| Indjabinda | Pilbara | WA |
| Jajaurung[20] | Northeast Region | VIC |
| Jauidjabaija | Yawajaba Island, Kimberleys | WA |
| Jindjiparndi[21] | Chichester Region, Pilbara | WA |
| Jiwarli | Henry River, North-western Region | WA |
| Jupagalk | Northwest Region | VIC |
| Kaitish | Barrow Creek, North of Alice Springs | NT |
| Kamilaroi | North Central Region | NSW |
| Karadjeri[22] | Lagrange Bay, Kimberleys | WA |
| Kaurna | Adelaide Plains | SA |
| Koko Ya'o[23] | Eastern Cape York | QLD |
| Kukata | South-Central Region | SA |
| Kukatja | Balgo, Lake Gregory, South Halls Creek | WA |
| Kuku-Yalangi* | Bloomfield River region | QLD |
| Kula | North-Central NSW, South-Central QLD | NSW/QLD |
| Kungarakany | Northwest Region, south of Darwin | NT |
| Kunwinjku | Arnhem Land | NT |
| Kurnai | Gippsland Region | VIC/NSW |
| Kurulk | Arnhem Land | NT |
| Kwadji | Eastern Cape York | QLD |
| Lardil | Mornington Island, Gulf of Carpentaria | QLD |
| Luritja[24] | Southeast of Alice Springs | NT |
| Mal Mal | Cleveland Bay near Townsville | QLD |
| Mara | Limmen Region, Arnhem Land | NT |
| Marduthunira | Cape Preston, Dampier, Pilbara Region | WA |
| Meru | Lower Murray River District | SA |
| Moporr | Between Leigh and Glenelg Rivers | VIC |
| Mukjarawaint | North-Central Region | VIC |
| Muruwari | North-Central NSW, South-Central QLD | NSW/QLD |
| Mycoolon | Saxby River Region | QLD |
| Nawu | Port Lincoln Area | SA |
| Narangga | Yorke Peninsula | SA |

---

[19] Dhauwurd Wurrung language, part of the Gundidjmara people
[20] aka Jaara, a subgroup of the Djadjawrung
[21] aka Yindjibarndi
[22] aka Karajarri
[23] Subgroup of the Kwadji
[24] aka Jumu

| | | |
|---|---|---|
| Narran | North-central region | NSW |
| Needwonee | Southwest Region | TAS |
| Ngalia | Leonora, Central Region | WA |
| Ngarigo | South-Central NSW, SE of Canberra | ACT/NSW |
| Ngemba | North-Central NSW, South-Central QLD | NSW/QLD |
| Ngarinman | Jasper Creek, Gregory National Park | NT |
| Ngarinyin | Kimberleys | WA |
| Ngarrindjeri | Lower Murray River District | SA |
| Noongahburrah [25] | Narran River, North-Central NSW | NSW |
| Nuenonne | Bruny Island, South-Central Region | TAS |
| Nungubuyu [26] | SE Arnhem Land, bordering Gulf of Carp. | NT |
| Plangermairrener | Ben Lomond District | TAS |
| Ramindjeri | Encounter Bay | SA |
| Ritarungo | Arnhem Land | NT |
| Tangani [27] | Lower Murray River District | SA |
| Tanganekald | Banks of the lower Murray River | SA |
| Tiwi | Bathurst & Melville Islands | NT |
| Turrbal [28] | Brisbane Area | QLD |
| Walangeri | (see Yarralin) | NT |
| Warai | Northwest Region, south of Darwin | NT |
| Wardaman | Upper Daly River District | NT |
| Warkawarka | Lake Tyrell Region | VIC |
| Warlpiri | Tanami Desert | NT |
| Wathi-Wathi | Murray River Region | VIC |
| Wawilak [29] | Arnhem Land | NT |
| Weilan | North-central region | NSW |
| Wheelman | Near Bremer Bay | WA |
| Wik-Munkan | Cape York Peninsula | QLD |
| Wiradjuri | Central Region | NSW |
| Wirangu | West coast | SA |
| Wolmeri | Kimberleys | WA |
| Wongaibon | North-central region | NSW |
| Worora | North of Derby, Kimberleys | WA |
| Wotjobaluk | Lake Albacutya Region | VIC |
| Wunambal | Kimberleys | WA |
| Yarralin | Victoria River District | NT |
| Yarripilangu | Area NE of Alice Springs | NT |
| Yarrungkanyi | Area NE of Alice Springs | NT |
| Yerrunthully | Central Region | QLD |
| Yiiji* | Forrest River Region, Kimberleys | WA |
| Yintjingga | Stewart River Area | WA |
| Yugerra* | Brisbane Region | QLD |

---

[25] Subgroup of the Narran
[26] This is the language of the Numbulwar of Arnhem Land
[27] Subgroup of the Ngarrindjeri
[28] aka Jagara, of the Yugerra Language group
[29] A clan of the Ritarungo People

## Appendix B: Evidence for a tsunami in the Gulf of Carpentaria

Evidence that tsunamis that struck the coastal regions of the Gulf of Carpentaria, including Groote Eylandt, Vanderlin Island, and parts of the eastern Northern Territory coast, has been proposed in the form of "chevron dunes", which Abbott et al (2007) and Tester and Abbott (2007) argue were formed when submarine sediment and coastal debris washed inland by the tsunami.  The chevrons all appear to strike from the southeast, in the direction of Kanmare structure in Figure 6.)

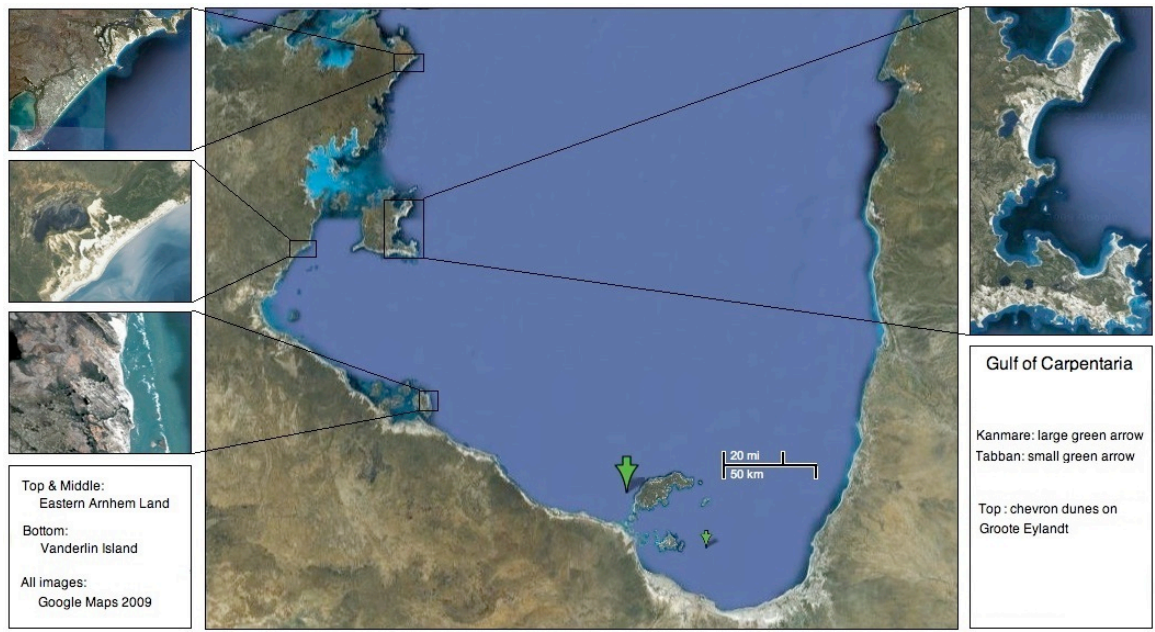

**Figure 6:** Chevron dunes on the coastal regions of the Gulf of Carpentaria, including the southeast coast of Groote Eylandt, the eastern coast of Arnhem Land, and Vanderlin Island, which strike from the southeast, towards the Kanmare structure.  A debate regarding the origins of these chevrons is ongoing.

## References

Abbott, Dallas H., Andrew Matzen, Edward A. Bryant, and Stephen F. Pekar
2003    Did a bolide impact cause catastrophic tsunamis in Australia and New Zealand? *Geological Society of America Abstracts with Programs* 35: 168

Abbott, Dallas H., Pierre E. Biscaye, Jihong Cole-Dai, and Dee Breger
2005a Evidence from an Ice Core of a Large Impact Circa 1443 A.D., *American Geophysical Union*, Fall Meeting 2005, abstract #PP31C-05

Abbott, Dallas H.; Suzanne Martos; Hannah Elkinton; Edward Bryant; Viacheslav Gusiakov and Dee Breger
2006    Impact Craters as Sources of Megatsunami Generated Chevron Dunes. Geological Society of America 2006 Philadelphia Annual Meeting, 22-25 October, 2006, Paper No.119-20, *Geological Society of America Abstracts with Programs* 38(7): 299

Abbott, Dallas H., Edward W. Tester, Cassaundra A. Meyers and Dee Breger
2007    Impact ejecta and megatsunami deposits from a historical impact into the Gulf of Carpentaria. *Geological Society of America Abstracts with Programs* 39(6): 312

Abbott, Dallas. H.; P. Gerard-Little; S. Costa and Dee Breger
2009    *Odd CaCO3 from the Southwest Indian Ocean Near Burckle Crater Candidate: Impact Ejecta or Hydrothermal Precipitate?* 40[th] Lunar and Planetary Science Conference (Lunar and Planetary Science XL), held March 23-27, 2009 in The Woodlands, Texas, id. 2243

Abbott, Dallas H.; Edward Bryant; Viacheslav Gusiakov and Bruce Masse
2010    Largest natural catastrophes in Holocene and their possible connection with comet-asteroid impacts on the Earth. *6th Alexander von Humboldt International Conference – Climate Change, Natural Hazards, and Societies*, held in Merida, Mexico on 14 – 19 March 2010, AvH6-13

Adcock, Gregory J., Elizabeth S. Dennis, Simon Easteal, Gavin A. Huttley, Lars S. Jermiin, W. James Peacock and Alan Thorne
2001    Mitochondrial DNA sequences in ancient Australians: Implications for modern human origins. *Proceedings of the National Academy of Sciences of the U.S.A.* 98 (2): 537–542

Alderman, Arthur Richard
1931    The meteorite craters at Henbury, Central Australia with an addendum by L. J. Spencer. *Mineralogical Magazine* 23:19-32
1936    The Carraweena, Yandama and Cartoonkana meteoritic stones. *Records of the South Australian Museum* 5: 537-546

Allen, Louis A.
1975    *Time Before Morning: art and myth of the Australian Aborigines.* Thomas Y. Crowell Co., New York

Anon
1880    Tenham fall. *The Western Champion*, Blackall, QLD, 19 May 1880, p. 3b
1870    An Aboriginal Legend. *The Sydney Morning Herald*, 08 August 1870, p. 4f
        http://nla.gov.au/nla.news-article13221055

Arunotai, Narumon
2006    Moken traditional knowledge: an unrecognised form of natural resources management and conservation. *International Social Science Journal* 58(187):139-150

Austin-Broos, Diane
1994    Narratives of the encounter at Ntaria. *Oceania* 65: 131-150
2009    *Arrernte Present, Arrernte Past: Invasion, Violence, and Imagination in*

*Indigenous Central Australia*. University of Chicago Press

Bailey, M. E., D. J. Markham, S. Massai and J.E. Scriven
1995   The 1930 August 13 'Brazilian Tunguska' event. *The Observatory* 115: 250-253

Baillie, Michael
2007   The case for significant numbers of extraterrestrial impacts through the late Holocene. *Journal of Quaternary Science* 22(2): 101-109

Baker, George
1957   The role of Australites in Aboriginal customs. *Memoirs of the National Museum of Victoria* 22(8): 1-26

Barker, Herbert M.
1964   *Camels and the Outback*. Sir Isaac Pitman and Sons Ltd., Melbourne

Bates, Daisy
1924   *Sky Stones*. Sydney Morning Herald, Saturday, 16 August 1924, p. 11d[30]
1996   *Aboriginal Perth and Bibbulmun biographies and legends*. Hesperian Press, Perth

Becker, L., R.J. Poreda, A.R. Basu, K.O. Pope, T.M. Harrison, C. Nicholson and R. Iasky
2004   Bedout: A Possible End-Permian Impact Crater Offshore of Northwestern Australia. *Science* 304(5676): 1469–1476

Beckett, Jeremy
1994   Aboriginal histories, Aboriginal myths: an introduction. *Oceania* 65: 97-115

Berndt, Ronald Murray and Catherine H. Berndt
1989   *The Speaking Land: myth and story in Aboriginal Australia*. Penguin Books Australia, Ltd.

Bevan, Alex W. R. and Peter Bindon
1996   Australian Aborigines and Meteorites. *Records of the Western Australian Museum* 18: 93-101

Bevan, Alex W. R. and R. A. Binns
1989   Meteorites for the Nullarbor region, Western Australia: I. A review of past recoveries and a procedure for naming new finds. *Meteoritics* 24: 127-133

Bevan, Alex W. R. and John de Laeter
2002   *Meteorites: a journey through space and time*. University of New South Wales Press, Ltd.

Bevan, Alex W. R. and Ken McNamara

---

[30] http://newspapers.nla.gov.au/ndp/del/article/16163145

1993    *Australia's meteorite craters*. Western Australian Museum, Perth

Bird, M.I.; C.S.M. Turney; L.K. Fifield; R. Jones; L.K. Ayliffe; A. Palmer; R. Cresswell and S. Robertson
2002    Radiocarbon analysis of the early archaeological site of Nauwalabila I, Arnhem Land, Australia: implications for sample suitability and stratigraphic integrity. *Quaternary Science Reviews* 21(8): 1061-1075

Blong, Russell J.
1982    *The time of darkness: local legends and volcanic reality in Papua New Guinea*. Australian National University Press, Canberra

Bobrowsky, Peter T. and Hans Rickman (Eds.)
2007    *Comet/Asteroid Impacts and Human Society*.  Springer Publishers

Boslough, M.B.E. and D.A. Crawford
2008    Low-altitude airbursts and the impact threat.  *International Journal of Impact Engineering* 35(12): 1441-1448

Bourgeois, Joanne and Robert Weiss
2009    "Chevrons" are not mega-tsunami deposits—A sedimentologic assessment. *Geology* 37(5): 403-406

Bowler, James M., Harvey Johnston, Jon M. Olley, John R. Prescott, Richard G. Roberts, Wilfred Shawcross and Nigel A. Spooner
2003    New ages for human occupation and climatic change at Lake Mungo, Australia. *Nature* 421: 837-840

Brown, Peter Lancaster
1975    *Comets, Meteorites, and Men*. The Scientific Book Club, London

Brown, P., R.E. Spalding, D.O. ReVelle, E. Tagliaferri, S.P. Worden
2002    The flux of small near-Earth objects colliding with the Earth.  *Nature* 420(6913): 294-296

Brown, Michael Fobes
2004    *Who owns native culture?* Harvard University Press

Bryant, Edward A.
2001    *Tsunami: the underrated hazard*. Cambridge University Press

Bryant, Edward A., Graham Walsh and Dallas H. Abbott
2007    Cosmogenic mega-tsunami in the Australia region: are they supported by Aboriginal and Maori legends?" In *Myth and Geology*, edited by L. Piccardi and W.B. Masse, Geological Society, London, Special Publications, 273

Bunch, Ted P.; James Kennett and Douglas K. Kennett
2008    Comment: Impacts, mega-tsunami, and other extraordinary claims. *GSA Today* 18(6): 11

Cassidy, W.A.
1954    The Wolf Creek, Western Australia, Meteorite Crater (CN=-1278,192). *Meteoritics* 1(2): 197

Cauchi, Stephen
2003    Outer space meets outback and a bluff is born. *The Age* 18 June 2003, http://www.theage.com.au/articles/2003/06/17/1055828328658.html

Charlesworth, Max; Howard Murphy; Francoise Dussart (Eds)
1984    *Religion In Aboriginal Australia: An Anthology of recent writings*.  University of Queensland Press

Chown, Marcus
1995    Did Falling Comet Cause Rumble in the Jungle?  *New Scientist* 148(2003): 12 (11 November 1995)

Clarkson, Jodi
1999    Consulting with central Australian Aboriginal people about cultural interpretation. In *The Human Factor in interpretation - Proceedings of the Interpretation Australia Association National Conference* held in Hobart, Tasmania in September 1999, pp. 30-36. Interpretation Australia Association, Collingwood, VIC.

Collins, Gareth S., H. Jay Melosh, and Robert A. Marcus
2005    Earth Impact Effects Program: A Web-based computer program for calculating the regional environmental consequences of a meteoroid impact on Earth. *Meteoritics and Planetary Science* 40(6): 817–840

Coon, Carlton S.
1972    *The Hunting Peoples*. Jonathan Cape, Ltd., London

Cowan, James
1992    *Aborigine Dreaming: an introduction to the wisdom and though of the Aboriginal traditions of Australia*. Thorson's Publishers, London

Dean, Colin
1996    *The Australian Aboriginal 'Dreamtime': Its History, Cosmogenesis Cosmology and Ontology*.  Gamahucher Press, West Geelong, Victoria

de Laeter, J.R.
1982    Two new iron meteorites from western Australia. *Meteoritics* 17(3): 135-140

Dietz, Robert S.
1967    Shatter Cone Orientation at Gosses Bluff Astrobleme. *Nature* 216(5120): 1082–1084

Dixon, Robert M. W.
1972    *The Dyirbal Language of North Queensland*. Cambridge University Press

Durda, Daniel D. and David A. Kring
2004    Ignition threshold for impact-generated fires.  *Journal of Geophysical Research* 109(E08004): E08004.1-E08004.14

Earth Impact Database
2009    Earth Impact Database <http://www.unb.ca/passc/ImpactDatabase/> (May 2009). Developed and maintained by the Planetary and Space Science Centre, University of New Brunswick, Canada.

Education Department of South Australia
1992    *The Adnyamathanha People.* Hyde Park Press, South Australia

Edwards, Robert
1966    Australites used for Aboriginal implements in South Australia.  *Records of the South Australian Museum* 15: 243-251

Elkinton, H.; Dallas Abbott; S. Martos and D. Breger
2006    Impactor Fragments from the Craters Kanmare and Tabban in the Gulf of Carpentaria, *Geological Society of America Abstracts with Programs* 38: 299

Firestone R.; Alan West and S. Warwick-Smith
2006    *The Cycle of Cosmic Catastrophes: Flood, Fire, and Famine in the History of Civilization*.  Bear & Company, Rochester, VT

Firestone, Richard B. and Alan West
2008    Comment: Impacts, mega-tsunami, and other extraordinary claims. *GSA Today* 18(6): e13

Fitzgerald, Michael John
1979    *The chemistry and mineralogy of meteorites from South Australia and adjacent regions*. PhD Thesis, Department of Geology and Mineralogy, University of Adelaide

Gallant, A. Roy
1996    Sikhote Alin Revisited.  *Meteorite Magazine* 2: 8-11 (February)

Gibbons, George Studley
1977    Index of Australian meteorites. *Journal of the geological Society of Australia* 24(5): 263-268

Gillespie, Richard and Richard G. Roberts
2000    On the reliability of age estimates for human remains at Lake Mungo. *Journal of Human Evolution* 38: 727-732

Glasstone, Samuel and Philip J. Dolan
1977    *The effects of nuclear weapons, 3rd edition*. United States Department of Defense and Department of Energy, Washington D.C.

Glikson, A. Y.; A.H. Hickman and J. Vickers
2008    Hickman Crater, Ophthalmia Range, Western Australia: evidence supporting a meteorite impact origin. *Australian Journal of Earth Sciences* 55(8): 1107-1117

Goff, James; Keri Hulme and Bruce McFadgen
2003    "Mystic Fires of Tamaatea": Attempts to creatively rewrite New Zealand's cultural and tectonic past. *Journal of the Royal Society of New Zealand* 33(4): 795-809

Goldsmith, John
2000    Cosmic impacts in the Kimberly. *Landscope Magazine* 15(3): 28-34

Green, Bill
2008    Shock discovery of huge meteorite crater. *Talking Headlines*, November 2008 http://www.talkingheadlines.com/shock-meteorite-discovery/

Gritzner, Christian
1997    Human Casualties in Impact Events. *WGN - Journal of the International Meteor Organization* 25(5): 222-226

Gusiakov, Viacheslav; Dallas Abbott; Edward Bryant; Bruce Masse
2008    Mega tsunami of the world ocean as the evidence of recent oceanic bolide impacts, chevron dune formation and rapid climate change. *Proceedings from the 33rd IGC International Geological Congress* held in Oslo, Norway on 6-14 August 2008

Haberle, Simon G.
2005    A 23,000-yr pollen record from Lake Euramoo, Wet Tropics of NE Queensland, Australia. *Quaternary Research* 64: 343-356

Haddon, Alfred C. (Edt)
1904    *Reports of the Cambridge Anthropological Expedition to Torres Straits*, Vol. V: *Sociology, Magic and Religion of the Western Islanders*. Cambridge University Press

Haines, Peter W.
2005    Impact Cratering and Distal Ejecta: The Australian Record. *Australian Journal of*

*Earth Sciences* 52: 481-507

Halliday, Ian; Alan T. Blackwell and Arthur A. Griffin
1984    The frequency of meteorite falls on the earth. *Science* 223(4643): 1405-1407

Hamacher, Duane W. and Ray P. Norris
2010a   Comets in Australian Aboriginal Astronomy. *Journal for Astronomical History and Heritage* (accepted for Nov 2010, in press)
2010b   Meteors in Australian Aboriginal Dreamings. *WGN – Journal of the International Meteor Organization* 38(3): 87-98

Harney, William Edward and Adolphus Peter Elkin
1949    *Songs of the Songmen: Aboriginal myths retold*. F.W. Cheshire, Melbourne

Haynes, Roslynn D.
2000    Astronomy and the Dreaming: the astronomy of the Aboriginal Australians. In *Astronomy Across Cultures: The History of Non-Western Astronomy*, edited by H. Selin and X. Sun, Dordrecht, Kluwer Academic Publishers

Hellyer, Brian
1971    The mass distribution of an aerolite shower (Tenham, Queensland, 1879). *The Observatory* 91: 64-66

Hill, D. H.; W. V. Boynton, and R. A. Haag
1991    A Lunar Meteorite Found outside the Antarctic. *Nature* 352(6336): 614–617

Hodge-Smith, Thomas
1939    Australian Meteorites. *Memoirs of the Australian Museum* 7: 1-84

Howard, Kieren T. and Peter W. Haines
2007    The geology of Darwin Crater, western Tasmania, Australia. *Earth and Planetary Science Letters* 260(1-2): 328-339

Huyghe, P.
1996    Incident at Curuçá. *The Sciences* 36(2): 14-17 (March/April)

Ivanov, Boris
2008    *Size-Frequency Distribution Of Asteroids And Impact Craters: Estimates Of Impact Rate*. In "*Catastrophic Events Caused by Cosmic Objects*", edited by Vitaly Adushkin and Ivan Nemchinov, Springer Publishers, Dordrecht, The Netherlands

Jankaew, Kruawun; Brian F. Atwater; Yuki Sawai; Montri Choowong; Thasinee Charoentitrat; Maria E. Martin and Amy Prendergast
2008    Medieval forewarning of the 2004 Indian Ocean Tsunami in Thailand. *Nature* 455: 1228-1231

Johnson, J. E. and D. H. McColl
1967    An aerolite from Cockburn, South Australia. *Transactions of the Royal Society of South Australia* 91: 37-40

Jones, Elise
1989    *The Story of the Falling Star.* Aboriginal Studies Press, Canberra

Jones, Tim P. and Bo Lim
2000    Extraterrestrial Impacts and Wildfires. *Palaeogeography, Palaeoclimatology, Paleoecology* 164(1–4): 57–66

Kaberry, Phyllis M.
1939    *Aboriginal woman: sacred and profane.* George Routledge and Sons, Ltd., London

Kenkmann, T. and M.H. Poelchau
2008    *Matt Wilson: an elliptical impact crater in Northern Territory, Australia.* 39[th] Lunar and Planetary Science Conference, Lunar and Planetary Science XXXIX, 10-14 March 2008, League City, Texas. LPI Contribution No. 1391, p. 1027 http://www.lpi.usra.edu/meetings/lpsc2008/pdf/1027.pdf

Kershaw, A.P.
1970    A Pollen Diagram from Lake Euramoo, North-East Queensland, Australia. *New Phytologist* 69(3): 785-805 (July)

Kohman, Truman P. and Parmatma S. Goel
1963    *Terrestrial ages of meteorites from cosmogenic 14C.* In "Radioactive dating", proceedings of the Symposium on Radioactive Dating held by the International Atomic Energy Agency in co-operation with the Joint Commission on Applied Radioactivity (ICSU) in Athens, Greece on 19-23 November 1962, Volume 1962, p. 395-411

Kruse P.D.; R.C. Maier; M. Khan and J.N. Dunster
2010    *Walhallow-Brunette Downs-Alroy-Frew River, Northern Territory. 1:250 000 geological map series explanatory notes, SE 53-07, SE 53-11, SE 53-15, SF 53-03.* Northern Territory Geological Survey, Darwin

Lewis, John S.
1999    *Comet and asteroid impact hazards on a populated earth: computer modeling.* Academic Press, San Diego/London

Lucich, Peter
1969    *Children's stories from the Worora.* Australian Aboriginal Studies No. 18, Social Anthropology Series No. 3, Australian Institute of Aboriginal Studies, Canberra

Macdonald, F.A. and K. Mitchell
2004    New possible, probable, and proven impact sites in Australia. *Geological Society of Australia Abstracts* 73: 239

Macey, Richard
2008    Opal miner stumbles on mega meteorite crater. *The Age* (Fairfax Digital) 23 November 2008, http://www.theage.com.au/news/technology/biztech/opal-miner-stumbles-on-mega-meteorite-crater/2008/11/22/1226770814042.html

Maddock, Kenneth
1982    *The Australian Aborigines: a portrait of their society*. Penguin Books Australia, Ringwood, Victoria

Madigan, Cecil Thomas
1937    The Boxhole Crater and the Huckitta Meteorite (Central Australia). *Royal Society of South Australia Transactions and Proceedings* 61:187-190

Madigan, Cecil Thomas and Arthur Richard Alderman
1939    The Huckitta Meteorite, Central Australia. *Mineralogical Magazine* 25(165): 353-371

Malbunka, Mavis
2009    *Tnorala*. Message Stick, Australian Broadcasting Company (ABC) Television. First aired on Sunday, 19 July 2009 at 13:00 on ABC-1. Transcripts and video available at: http://www.abc.net.au/tv/messagestick/stories/s2629034.htm

Martos, Suzanne N., Dallas H. Abbott, Hannah D. Elkinton, Allan R. Chivas and Dee Breger
2006    Impact spherules from the craters Kanmare and Tabban in the Gulf of Carpentaria. *Geological Society of America Abstracts with Programs* 38(7): 299

Masse, W. Bruce and Michael J. Masse
2007    Myth and catastrophic reality: using myth to identify cosmic impacts and massive Plinian eruptions in Holocene South America. In *Myth and Geology*, edited by L. Piccardi and W.B. Masse, Geological Society, London, Special Publications, 273

Masse, W. Bruce, Elizabeth Wayland Barber, Luigi Piccardi and Paul T. Barber
2007a   Exploring the nature of myth and its role in science. In *Myth and Geology*, edited by L. Piccardi and W.B. Masse, Geological Society, London, Special Publications 273

Masse, W. Bruce, R.P. Weaver, Dallas Abbott, V.K. Gusiakov and Edward Bryant
2007b   *Missing in action? Evaluating the putative absence of impacts by large asteroids and comets during the Quaternary Period*. Proceedings of the Advanced Maui Optical and Space Surveillance Technologies conference, Wailea, Maui, Hawaii, pp. 701-710

Master, Sharad and T. Woldai
2004    The Umm al Binni structure in the Mesopotamian marshlands of southern Iraq, as a postulated late Holocene meteorite impact crater: geological setting and new Landsat ETM and Aster satellite imagery. *Economic Geology Research Institute Information Circular*, October 2004, University of Witwatersrand - Johannesburg, South Africa

Mathews, Janet
1994    *The opal that turned to fire and other stories from the Wangkumara*. Magabala Books Aboriginal Corporation, Broome, Western Australia

McCall, Gerald Joseph Home and John R. de Laeter
1965    *Catalogue of Western Australian Meteorite Collections*.   Special publication, Western Australian Museum 3

McKay, Helen F., Pauline E. McLeod, Francis Firebrace Jones and June E. Barker
2001    *Gadi Mirrabooka: Australian Aboriginal tales from the Dreaming*. Libraries Unlimited, Greenwood Pub Group, Inc., Englewood, Colorado

McNiven, Ian J., Bruno David, Goemulgau Kod and Judith Fitzpatrick
2009    The Great *Kod* of Pulu: Mutual Historical Emergence of Ceremonial Sites and Social Groups in Torres Strait, Northeast Australia. *Cambridge Archaeological Journal* 19(3): 291–317

Meggitt, M.J.
1972    Understanding Australian Aboriginal society: kinship systems or cultural categories? In *Kinship Studies in the Morgan Centennial Year*, edited by P. Reining, Anthropological Society of Washington D.C., pp. 64-87

Michaels, Eric
1985    Constraints on Knowledge in an Economy of Oral Information.   *Current Anthropology* 26(4): 505-510

Milton, Daniel J., Andrew Y. Glikson and Robin Brett
1996    Gosses Bluff—a latest Jurassic impact structure, central Australia. Part 1: geological structure, stratigraphy, and origin. *AGSO Journal of Australian Geology and Geophysics* 16(4): 453–486

Morrison, David; Alan W. Harris; Geoff Sommer; Clark R. Chapman and Andrea Carusi
2003    Dealing with the Impact Hazard.   In '*Asteroids III*' edited by William Bottke, Alberto Cellino, Paolo Paolicchi, and Richard P. Binzel, University of Arizona Press

Mountford, Charles Pearcy
1976    *Nomads of the Australian Desert*. Rigby, Ltd., Adelaide

Mowaljarlai, David and Jutta Malnic
1993    *Yorro Yorro: Aboriginal creation and the renewal of nature: rock paintings and stories from the Australian Kimberley.* Magabala Books Aboriginal Corporation, Broome, Western Australia

Murgatroyd, W.
2001    Grace Resources NL, Magnesium Mine, Batchelor, NT: environmental impact assessment, anthropological component, initial report.   A report to URS Environmental Consultants, Darwin, NT.  (See also  C.W. MacPherson Collection, Photo #PH0346/0162, Northern Territory Library,

Napier, William M. and David Asher
2009    The Tunguska impact event and beyond.  *Astronomy & Geophysics* 50(1): 1.18-1.26

Napier, William M.; J.T. Wickramasinghe and N.C. Wickramasinghe
2004    Extreme albedo comets and the impact hazard. *Monthly Notices of the Royal Astronomical Society* 355: 191-195

Neate, Graeme
1982    Keeping Secrets Secret: Legal Protection for Secret/Sacred items of Aboriginal Culture.  *Aboriginal Law Bulletin* 1(5): 1
        http://www.austlii.edu.au/au/journals/AboriginalLB/1982/42.html

O'Connell, Edna
1965    *A catalog of meteorite craters and related features with a guide to the literature*. Rand Corp.

O'Connell, James F. and Jim Allen
2004    Dating the colonization of Sahul (Pleistocene Australia–New Guinea): A review of recent research.  *Journal of Archaeological Science* 31: 835-853

Palmer, Edward
1884    Notes on Some Australian Tribes.  *The Journal of the Anthropological Institute of Great Britain and Ireland* 13: 276-347

Pannell, Sandra
2006    *Reconciling Nature and Culture in a Global Context: Lessons form the World Heritage List.* Cooperative Research Centre for Tropical Rainforest Ecology and Management, James Cook University, Cairns, QLD, Australia
        http://www.jcu.edu.au/rainforest/publications/nature_culture_text.pdf

Parks and Wildlife Commission of the Northern Territory
1997    *Tnorala Conservation Reserve (Gosse's Bluff) Plan of Management*, March 1997, Alice Springs, Amended May 2007

http://nt.gov.au/nreta/parks/manage/plans/pdf/tnorala.pdf
2002    *Henbury Meteorites Conservation Reserve: Draft plan of management*, November 2002, http://www.nt.gov.au/nreta/parks/manage/plans/pdf/henbury_pom.pdf

Pascoe, Bruce (Edt)
1990    *Aboriginal Short Stories, No 32.*  Pascoe Publishing, Apollo Bay, VIC, Australia

Peck, Charles William
1925    *Australian Legends: tales handed down from the remotest times by the autochthonous inhabitants of our land, Parts I and II.* Stafford and Co. Ltd., Sydney
1933    *Australian Legends.* Lothian, Melbourne

Piccardi, L. and W. Bruce Masse (eds)
2007    *Myth and Geology.*  Geological Society, London, Special Publications, 273

Pinter, Nicholas and Scott E. Ishman
2008    Impacts, mega-tsunami, and other extraordinary claims. *GSA Today* 18(1): 37-38

Poirier, Sylvie
2005    *A World of Relationships: Itineraries, Dreams, and Events in the Australian Western Desert.* University of Toronto Press

Radcliffe-Brown, A.R.
1926    The Rainbow-Serpent Myth of Australia. *Journal of the Royal Anthropological Institute of Great Britain and Ireland* 56(1926): 19-25

Rainforest Conservation Society of Queensland
1986    *Tropical Rainforests of North Queensland: Their Conservation Significance.* Report to the Australian Heritage Commission by the Rainforest Conservation Society of Queensland, Special Australian Heritage Publication Series No. 3. Australian Government Publishing Service, Canberra

Rappenglück, B.; Ernstson, K.; Mayer, W.; Neumair, A.; Rappenglück, M.A.; Sudhaus, D. and Zeller, K.W.
2009    The Chiemgau Impact: An Extraordinary Case Study for the Question of Holocene Meteorite Impacts and their Cultural Implication. *Cosmology Across Cultures* 409: 338

Redd, A. J. and M. Stoneking
1999    Peopling of Sahul: mtDNA Variation in Aboriginal Australian and Papua New Guinean Populations. *The American Journal of Human Genetics* 65(3): 808-828

Reed, Alexander Wyclif
1993    *Aboriginal Myths, Legends and Fables.* Reed New Holland, London

Roberts, Ainslie and Charles P. Mountford
1965    *The Dreamtime: Australian Aboriginal myths in paintings by Ainslie Roberts with text by Charles Mountford*. Rigby Ltd., Adelaide

Roberts, R. G., R. F. Galbraith, J.M Olley, H. Yoshida, G. M. Laslett
1999    Optical dating of single and multiple grains of quartz from Jinmium Rock Shelter, northern Australia: Part II, results and implications. *Archaeometry* 41(2): 365-395

Robinson, Roland
1966    *Aboriginal myths and legends*. Sun Books Pty, Ltd., Melbourne

Róheim, Géza
1945    *The Eternal Ones of the Dram: a psychoanalytic interpretation of Australian myth and ritual*. International Universities Press, New York
1971    *Australian Totemism: a psycho-analytic study in anthropology*. Frank Cass and Co., Ltd., London

Rose, Deborah Bird
1996    *Nourishing terrains: Australian Aboriginal views of landscape and wilderness.* Australian Heritage Commission, Canberra
2000    *Dingo makes us human: life and land in an Australian aboriginal culture.* Cambridge University Press

Ross, Margaret Clunies
1986    Australian Aboriginal Oral Traditions. *Oral Tradition* (1/2): 231-271

Rumsey, Alan
1994    The Dreaming, human agency and inscriptive practice. *Oceania* 65: 116-130

Sanday, Peggy Reeves
2007    *Aboriginal Paintings of the Wolfe Creek Crater: Track of the Rainbow Serpent*, University of Pennsylvania Press

Sawtell, Michael
1955    The legend of the falling star. *Dawn: A Magazine for the Aboriginal People of New South Wales* 4(2): 21-22

Shoemaker, Eugene M., Carolyn S. Shoemaker, Kunihiko Nishiizumi, Candace P. Kohl, James R. Arnold, Jeffery Klein, David Fink, Roy Middleton, Peter W. Kubik and Pankaj Sharma
1990    Ages of Australian meteorite craters - a preliminary report. *Meteoritics* 25: 409

Shoemaker, Eugene M., F.A. MacDonald and Carolyn S. Shoemaker,
2005    Geology of five small Australian impact craters. *Australian Journal of Earth Sciences* 52: 529-544

Shoemaker, Carolyn S. and Francis A. MacDonald
2005   The Shoemaker legacy to the Australian impact record. *Australian Journal of Earth Sciences* 52: 477-479

Simpson, Edward S.
1938   Some new and little known meteorites found in Western Australia. *Mineralogical Magazine* 25: 157-171

Smith, M.A.
1987   Pleistocene occupation in arid Central Australia. *Nature* 328(6132): 710-711

Smith, M.A.; M.I. Bird; C.S.M. Turney; L.K. Fifield; G.M. Santos; P.A. Hausladen and M.L. di Tada
2001   New abox AMS $^{14}$C ages remove dating anomalies Puritjarra rock shelter. *Australian Archaeology 53*: 45-47

Smith, W. Ramsay
1970   *Myths and Legends of the Australian Aboriginals.* George G. Harrap and Co. Ltd., London

Spencer, L.J.
1937   Two new meteoric stones from South Australia - Lake Labyrinth and Kappakoola. *Mineralogical Magazine* 24: 353-361

Sweet, I.P.; A.T. Brakel; D.J. Rawlings; P.W. Haines; K.A. Plumb and A.S. Wygralak
1999   *Mount Marumba, Northern Territory 1:250,000 Geological Map Series Explanatory notes SD53-6, 2nd Edt*. Australian Geological Survey Organisation and Northern Territory Geological Survey, National Geoscience Mapping Accord

Stanner, William Edward Hanley
1958   The Dreaming.   In *Reader in Comparative Religion: an anthropological approach*, edited by W. A. Lessa and E. Z. Vogt, Harper and Row, New York
1965   Religion, totemism and symbolism. In *Aboriginal Man in Australia: Essays of Honour of Emeritus Professor A. P. Elkin*, edited by R. M. Berndt and C. H. Berndt, Angus and Robertson, Sydney
1975   *Australian Aboriginal Mythology: Essays in Honour of W. E. H. Stanner*, edited by Lester Richard Hiatt.   Australian Institute of Aboriginal Studies, Canberra. Social Anthropology Series 9

Steel, Duncan
1995   *Rogue Asteroids and Doomsday Comets: The search for the million-megaton menace that threatens life on Earth.* John Wiley and Sons, Hoboken, New Jersey
1996   A Tunguska event in British Guyana in 1935? *Meteorite Magazine* 2: 12-13 (February)

Steel, Duncan and Peter Snow
1992    The Tapanui region of New Zealand: Site of a Tunguska around 800 years ago? In *Asteroids, Comets, Meteors 1991,* edited by A. Harris and E. Bowell, Lunar and Planetary Institute, Houston, Texas

Tester, Edward W. and Dallas Abbott
2007    Evidence for late Holocene oblique impact in the Gulf of Carpentaria, Australia. *Geological Society of America Abstracts with Programs* 39(6): 312

Thompson, Donald F.
1933    The Hero Cult, Initiation and Totemism on Cape York.  *The Journal of the Royal Anthropological Institute of Great Britain and Ireland* 63: 453-537 (July – Dec)

Thorne, Alan, Rainer Grün, Graham Mortimer, Nigel A. Spooner, John J. Simpson, Malcolm McCulloch, Lois Taylor and Darren Curnoe
1999    Australia's oldest human remains: age of the Lake Mungo 3 skeleton. *Journal of Human Evolution* 36:591-612.

Threlkeld, Lancelot Edward
1834    *An Australian Grammar: comprehending the principles and natural rules of the language, as spoken by the Aborigines in the vicinity of Hunter's River, Lake Macquarie, &c. New South Wales.*  Stephens & Stokes, Sydney

Tindale, Norman Barnett
1938    Prupe and Koromarange: A legend of Tanganekald, Coorong, South Australia. *Transactions of the Royal Society of South Australia* 62(1): 18-25
1946    Australian (Aborigine).  In *Encyclopedia of Literature*, edited by J.T. Shipley, Philosophical Library, New York, pp. 74-78
2005    Celestial lore of some Australian tribes.  In *Songs from the Sky: Indigenous Astronomical and Cosmological Traditions of the World*, edited by Von Del Chamberlain, J. B. Carlson and M. J. Young. Ocarina Books, Ltd: Bognor Regis; Center for Archaeoastronomy, College Park, Maryland

Turbet, Peter
1989    *The Aborigines of the Sydney district before 1788.*  Kangaroo Press, Kenthurst NSW

Vitaliano, Dorothy B.
1968    Geomythology. *Journal of the Folklore Institute* 5(1): 5-30
2007    Geomythology: geological origins of myths and legends. In *Myth and Geology*, edited by L. Piccardi and W.B. Masse, Geological Society, London, Special Publications, 273

Voris, Harold K.
2001    Maps of Pleistocene sea levels in Southeast Asia: shorelines, river systems and time durations. *Journal of Biogeography* 27: 1153-1167

Walsh, Michael
1991    Overview of Indigenous languages of Australia.  In *Language in Australia*, edited by Suzanne Romaine, Cambridge University Press

Warlukurlangu Artists
1987    *Kuruwarri: Yuendumu doors*. Australian Institute of Aboriginal Studies, Canberra

Warner, W. Lloyd
1937    *A Black Civilization: A social study of an Australian Tribe*. Harper and Brothers Publishers, New York

Williams, Warren H.
2004    *Warren H. Williams, the stories, the songs, Clip 2: Falling from the sky*.  Central Australian Aboriginal Media Association (CAAMA) Productions, Alice Springs

Wlotzka, Frank (Edt)
1991    Meteoritical Bulletin, No. 71.  *Meteoritics* 26: 255-262

Worms, Ernest Ailred
1940    The Southward Wandering.  *Annali Lateranensi* 4(5): 45,78
1943/44 Aboriginal place names in Kimberley, Western Australia. *Oceania* 14: 284-310

Yau, Kevin, Paul Weissman and Donald Yeomans
1994    Meteorite Falls in China and Some Related Human Casualty Events.  *Meteoritics* 29: 864-871

Zolensky, M. E., G.L. Wells and H.M. Rendell
1990    The accumulation rate of meteorite falls at the earth's surface - The view from Roosevelt County, New Mexico.  *Meteoritics* 25: 11-17